\documentclass[aps,pra,floatfix,a12]{revtex4}
\usepackage{amsfonts}
\usepackage{amsmath}
\usepackage{amsthm}
\usepackage{amssymb}
\usepackage{graphicx}
\usepackage{appendix}
\usepackage{hyperref}
\usepackage{textcomp}
\usepackage{multirow}
\usepackage{color}
\usepackage{bm}
\usepackage{braket}
\usepackage{subfigure}

\begin{document}

\title{Two-dimensional solitons in nonlocal media: a brief review}
\author{Boris A. Malomed$^{1,2}$}
\address{$^{1}$Department of Physical Electronics, School of Electrical Engineering, Faculty of Engineering, and Center for Light-Matter Interaction, Tel Aviv University, Tel Aviv 69978, Israel\\
$^{2}$Instituto de Alta Investigaci\'{o}n, Universidad de Tarapac\'{a}, Casilla 7D, Arica, Chile}

\begin{abstract}
This is a review addressing soliton-like states in systems with nonlocal
nonlinearity. The work on this topic has long history in optics and related
areas. Some results produced by the work (such as solitons supported by
thermal nonlinearity in optical glasses, and orientational nonlinearity
which affects light propagation in liquid crystals) are well known, and have
been properly reviewed in the literature, therefore the respective models
are outlined in the present review in a brief form. Some other studies, such
as those addressing models with fractional diffraction, which is represented
by a linear nonlocal operator, have started more recently, therefore it will
be relevant to review them in detail when more results will be accumulated;
for this reason, the present article provides a short outline of the latter
topic. The main part of the article is a summary of results obtained for
two-dimensional solitons in specific nonlocal nonlinear models originating
in studies of Bose-Einstein condensates (BECs), which are sufficiently
mature but have not yet been reviewed previously (some results for
three-dimensional solitons are briefly mentioned too). These are, in
particular, \textit{anisotropic} quasi-2D solitons supported by long-range
dipole-dipole interactions in a condensate of magnetic atoms (Tikhonenkov,
Malomed, and Vardi, 2008a), and \textit{giant vortex solitons} (which are
stable for high values of the winding number: Qin, Dong, and Malomed
(2016)), as well as 2D vortex solitons of the latter type \textit{moving
with self-acceleration} (Qin \textit{et al}., 2019). The vortex solitons are
states of a hybrid type, which include matter-wave and electromagnetic-wave
components. They are supported, in a binary BEC composed of two different
atomic states, by the resonant interaction of the two-component matter waves
with a microwave field which couples the two atomic states. The shape,
stability, and dynamics of the solitons in such systems are strongly
affected by their symmetry. Some other topics are included in the review in
a brief form.

This review uses the \textquotedblleft Harvard style" of referring to the
bibliography.
\end{abstract}

\author{}
\maketitle
\tableofcontents

\section{Introduction}

The absolute majority of work which has been performed in the huge area of
theoretical and experimental studies of solitons dealt with one-dimensional
(1D) settings. Extension of the soliton concepts to the multidimensional
world is a very promising, but also really challenging, direction for the
work of theorists and experimentalists. The obvious gain offered by
considering 2D and 3D soliton physics is the possibility to create
completely new species of localized states -- in particular, because 2D and
3D geometries make it possible to build localized topological modes with
intrinsic vorticity. Multi-component solitons can be used to build more
sophisticated topological structures, such as famous \textit{skyrmions},
\textit{hopfions} (alias twisted vortex tori in the 3D space), knots, and
others, which have no 1D counterparts (Manton and Sutcliffe, 2004; Radu and
Volkov, 2008).

However, the work with solitons in the 2D and 3D geometries encounters
fundamental difficulties. First, the most fundamental models that give rise
to 1D solitons, such as the Korteweg - de Vries (KdV), sine-Gordon (SG), and
nonlinear Schr\"{o}dinger (NLS) equations, are integrable by means of the
inverse-scattering transform and related methods (Zakharov \textit{et al}.,
1980; Ablowitz and Segur, 1981; Newell, 1985). Both the SG and NLS equations
have straightforward 2D and 3D versions which, however, are not integrable
and 3D their 2D and 3D counterparts are not integrable. As concerns the KdV
equation, its natural 2D extension, \textit{viz}., the two celebrated
Kadomtsev-Petviashvili equations (KP-I and KP-II, which differ by the sign
of the 2D spatial-dispersion term), provide exceptional examples of
integrable 2D equations (see a review article by Biondini and Pelinovsky,
2008; the integrability of the KP equations and the existence of 2D solitons
produced by them was discovered by Manakov \textit{et al}., 1977). Actually,
the lack of the integrability of basic 2D and 3D equations in the soliton
theory is only a technical difficulty, because relevant solutions can be
readily constructed in the numerical form (Yang, 2010), and, quite often, by
means of approximate analytical methods, such as the ubiquitous variational
approximation (VA); for the first time, the VA was applied to 2D solitons of
the NLS equation by Desaix, Anderson and Lisak (1991).

A principal problem is that the exit from 1D models to the 2D and 3D world
leads to versatile \emph{instabilities}, which do not occur in 1D. The
problem is clearly exhibited by the NLS equation with the self-attractive
cubic nonlinearity, which represents the Kerr term in optics (Chiao,
Garmire, and Townes, 1964), or attractive inter-atomic interactions in
Bose-Einstein condensates (BECs; Bradley \textit{et al}., 1995): while
stationary soliton solutions of the 1D NLS equations are commonly known to
be completely stable, the 2D and 3D versions of the same equation produce
soliton families that are \emph{completely unstable}, due to the fact that
precisely the same NLS equation gives rise to the \emph{collapse}, alias
blowup, i.e., catastrophic self-compression of the wave field leading to the
formation of a true singularity after a finite evolution time (Berg\'{e},
1998; Sulem and Sulem, 1999; Fibich, 2015), as illustrated by Fig. \ref%
{fig1.15}(a). The collapse in \textit{critical} in 2D, and \textit{%
supercritical} in 3D, which means that the 2D collapse sets in if the norm
of the input exceeds a certain finite critical (threshold) value, while in
3D the threshold is zero, i.e., an arbitrarily weak input may initiate the
supercritical collapse. In 2D, the input whose norm falls below the
threshold value does not blow\ up, but instead decays into \textquotedblleft
radiation" (small-amplitude waves). Thus, small perturbations added to any
soliton of the 3D NLS equation trigger its blowup, while in 2D the addition
of small perturbations initiates either the blowup or decay. In this
connection, it is relevant to mention that the first species\emph{\ }of
solitons which was\emph{\ }ever considered in optics is the family of the
so-called \textit{Townes solitons} (TSs), predicted by Chiao, Garmire, and
Townes (1964), without the analysis of their stability. As shown in Fig. \ref%
{fig1.15}(b), these are stationary solutions of the 2D NLS equation which
predict self-trapped shapes of laser beams propagating in the bulk Kerr
medium, under the condition of paraxial diffraction. In the original work,
these beams were not called solitons, as this term was coined only the next
year by Zabusky and Kruskal (1965). Many other species of optical solitons,
that were predicted later, have been created in the experiment, but the TSs
in their pure form have never been observed in optics, as they are unstable
states which represent the separatrix between collapsing and decaying
solutions of the 2D NLS equation (recently, experimental observation of TSs,
at the pre-blowup stage, in the effectively two-dimensional self-attractive
BECs in an ultracold atomic gas was reported by Chen and Hung (2020, 2021)).

\begin{figure}[tbp]
\begin{center}
\subfigure[]{\includegraphics[width=0.45\textwidth]{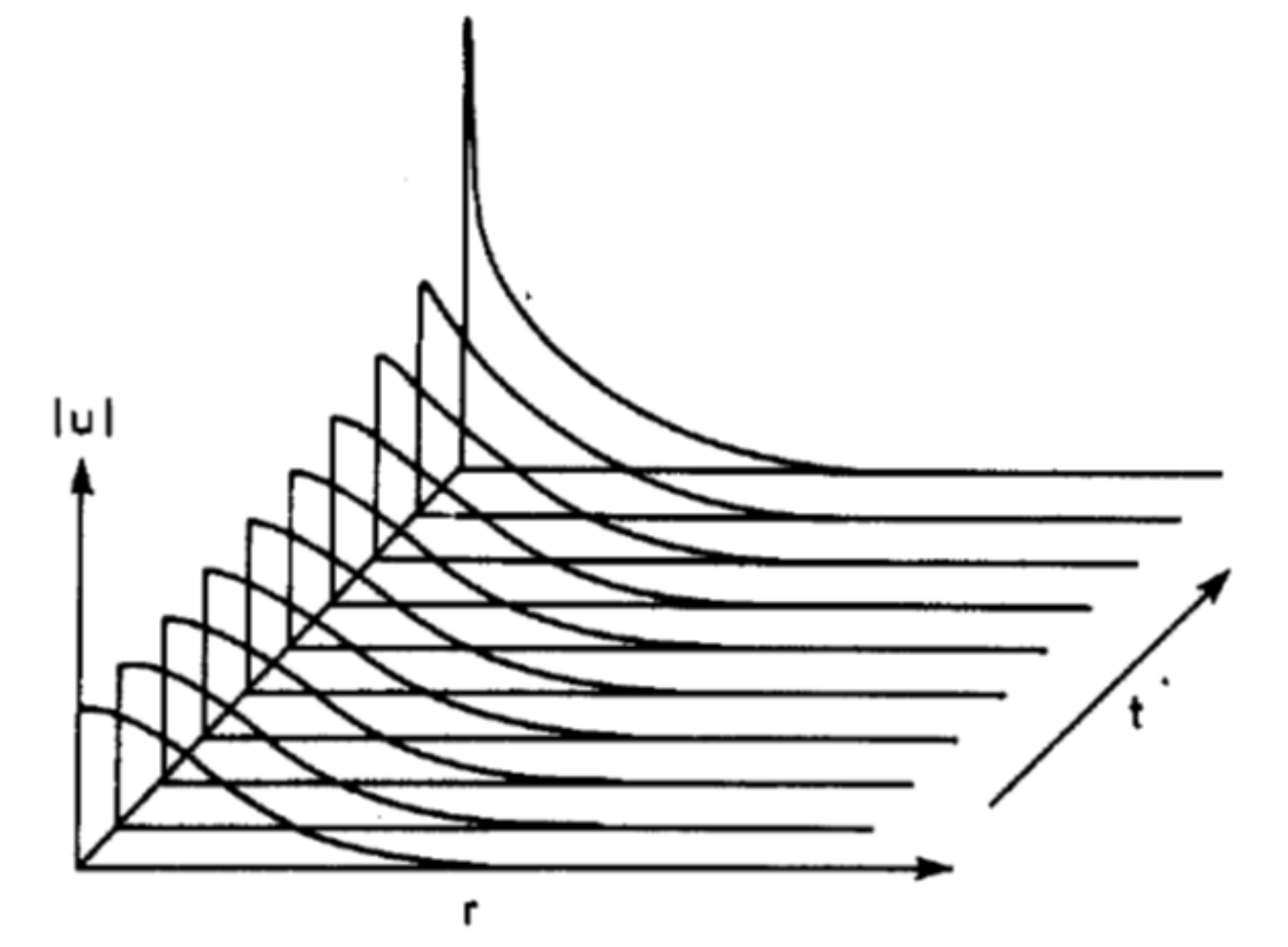}} %
\subfigure[]{\includegraphics[width=0.45\textwidth]{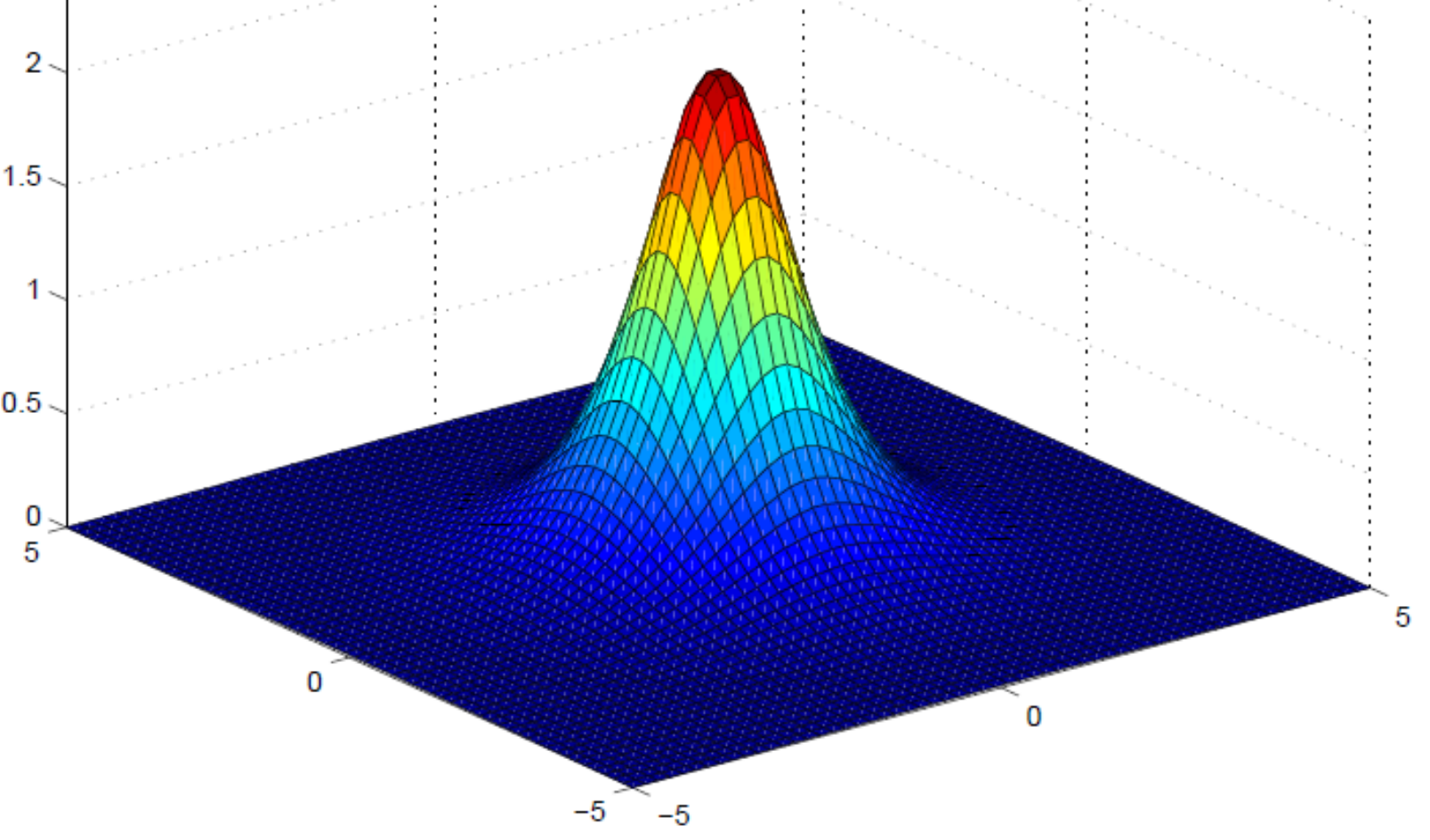}}
\end{center}
\caption{(a) Development of the collapse of the 2D Townes soliton, shown in
its radial cross-section. (b) The full spatial profile of the same soliton
in the stationary state (source:
https://www2.mathematik.uni-halle.de/dohnal/SOLIT\_WAVES/NLS\_blowup.pdf).}
\label{fig1.15}
\end{figure}

\begin{figure}[tbp]
\begin{center}
\includegraphics[width=0.65\textwidth]{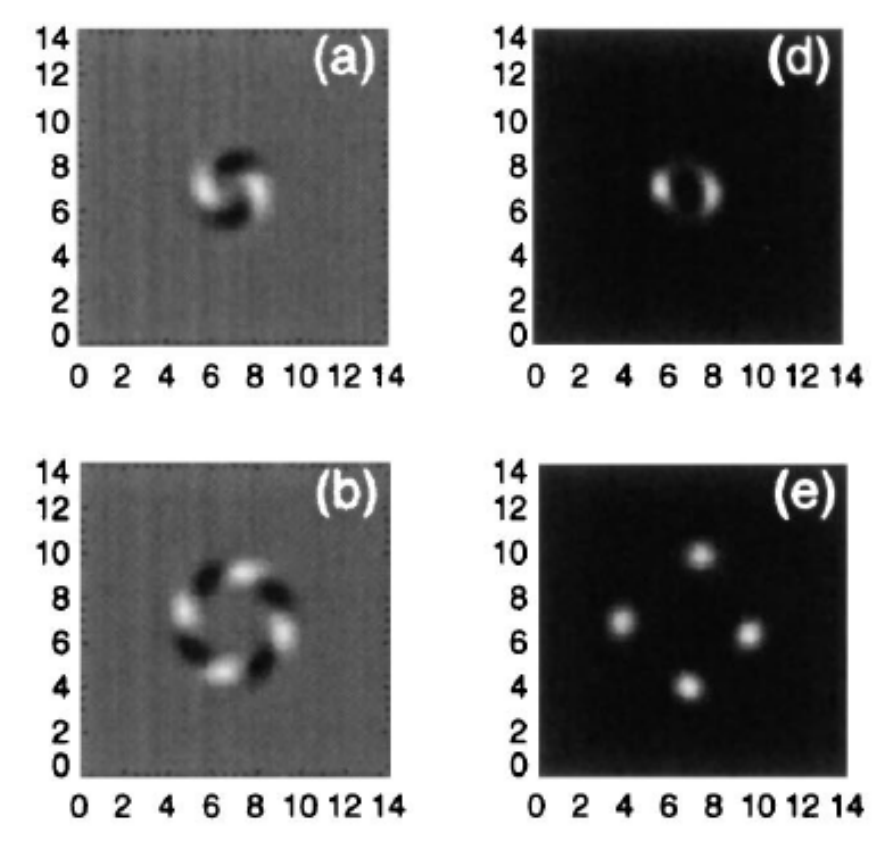}
\end{center}
\caption{Spontaneous splitting of unstable 2D VRs (vortex rings) with
winding numbers $S=1$ (panels a,d) and $S=2$ (b,e) into fundamental
(zero-vorticity) solitons. Panels (a) and (b) display the real part of the
complex wave function of the initial VRs, while (d) and (e) show the
intensity distribution in the fragments produced by the splitting. These
results were produced by simulations of the 2D NLS equations with saturable,
rather than cubic, self-focusing nonlinearity, which stabilizes
zero-vorticity solitons against the collapse, but does not stabilize the VRs
against the splitting. Note that the total angular momentum is conserved, as
the spin momentum of the initial VR is transformed into the orbital momentum
of the emerging fragments (source: Firth and Skryabin, 1997).}
\label{fig:splitting}
\end{figure}

As concerns 2D and 3D solitons with embedded vorticity (alias \textit{vortex
rings}, VRs), they are subject to the annular modulational instability,
which develops faster than the collapse, leading to spontaneous splitting of
the VR into two or several fragments, that are close to the corresponding
fundamental (zero-vorticity) solitons. The exact number of the fragments is
determined by the integer winding number (vorticity) carried by the VR, as
shown in Fig. \ref{fig:splitting}. At a later stage of the evolution, the
secondary solitons are destroyed by the collapse. In particular,
vorticity-carrying varieties of the (unstable) 2D TSs were introduced in the
works by Kruglov and Vlasov (1985), Kruglov \textit{et al}. (1988), and
Kruglov, Logvin, and Volkov (1992).

While the NLS equation with the self-attractive nonlinearity is a relevant
model for many physical realizations in optics (Kivshar and Agrawal, 2003),
BEC (Pitaevskii and Stringari, 2003), physics of Langmuir waves in plasmas
(Schochat and Weinstein, 1986), \textit{etc}., the occurrence of the
collapse implies that these physical settings cannot be used for the
straightforward creation of multidimensional solitons. Therefore, a\
cardinal problem is search for physically realistic multidimensional systems
which include additional ingredients that make it possible to suppress the
collapse and help to predict and create stable (or maybe metastable) 2D and
3D solitons, see reviews by Malomed \textit{et al}. (2005 and 2016), Malomed
(2016 and 2019), and Kartashov \textit{et al}. (2019), and a new book by
Malomed (2022). This can be done in various physical setups. In particular,
stable 2D and 3D optical solitons can be predicted, and, eventually,
experimentally created, if the optical medium features, in addition to the
cubic self-focusing, higher-order quintic self-defocusing, that arrests the
blowup, and thus provides the stabilization of 2D and 3D optical solitons.
The creation of fundamental 2D solitons stabilized by the quintic
self-defocusing was reported in the experiment by Fal\~{c}ao Filho \textit{%
et al}. (2013), and transitionally stable vortex solitons in the same medium
were observed by Reyna \textit{et al}. (2016). On the other hand, creation
of stable 3D solitons remains a challenging problem.

Another extremely interesting option is to consider a binary BEC with the
collapse driven by the attractive cubic interaction between its two
intrinsically self-repulsive components. In this system, the collapse is
arrested by a higher-order quartic self-repulsive term, which is induced in
each component by the correction to the cubic mean-field interaction,
induced by quantum fluctuations (the effect first addressed in the classical
work by Lee, Huang, and Yang, 1957). As a result, the binary BEC creates
completely stable 3D and quasi-2D self-trapped \textquotedblleft quantum
droplets" (QDs), which seem as multidimensional solitons (even if \ they are
not usually called \textquotedblleft solitons", as the name of QDs is
preferred in the literature). The prediction of QDs by Petrov (2015) and
Petrov and Astrakharchik (2016) was quickly realized experimentally (Cabrera
\textit{et al}, 2018; Cheiney \textit{et al}., 2018; Semeghini \textit{et al}%
. 2018; Ferioli \textit{et al}., 2019; D'Errico \textit{et al}., 2019). The
existence of stable QDs with embedded vorticity was predicted too (Kartashov
\textit{et al}., 2018; Li \textit{et al}., 2018), but such donut-shaped
vortex tori have not yet been created in the experiment.

As concerns 2D and 3D settings with the purely cubic nonlinearity, it was
predicted that completely stable 2D solitons can be created in two-component
systems with the spin-orbit coupling (SOC)\ between the components
(Sakaguchi, Li, and Malomed, 2014; Sakaguchi, Sherman, and Malomed, 2016).
Moreover, SOC may also create metastable solitons (ones which are stable
against small perturbations, while the supercritical collapse remains
possible) in the full 3D version of the same two-component system (Zhang
\textit{et al}. 2015b). Due to the specific form of the SOC, both 2D and 3D
two-component solitons maintained by this linear interaction between the
components take the shape of \textit{semi-vortices}, i.e., complexes
including a zero-vorticity soliton in one component and a vortical one in
the other, or \textit{mixed modes}, in which zero-vorticity and
vorticity-carrying terms are mixed in both components. In addition to that,
stable 3D solitons may be supported by a combination of SOC, taken in the
reduced 2D form, with the Zeeman splitting between the two components of the
BEC (Kartashov \textit{et al}., 2020a). Stable complexes of coupled 2D
solitons can be supported by spatially periodic modulation of the local SOC
strength (Kartashov \textit{et al}., 2020b).

All the above-mentioned mechanisms provide stabilization of 2D and 3D
solitons in models with local nonlinear self- and cross-interactions in the
single- and multi-component systems, respectively. On the other hand, a
straightforward possibility is to use nonlocal nonlinearities for the
stabilization of multidimensional localized states. First of all, it is
evident that a fixed spatial scale (correlation length) of the nonlinear
interaction arrests the development of the collapse, preventing the creation
of the singularity with a vanishingly small intrinsic scale. Because the
onset of the collapse is the basic mechanism leading to the instability of
solitons in 2D and 3D spaces, the nonlocality may be a powerful method
providing for the stabilization of such solitons. The present article offers
a review of some selected results produced by the work performed in this
direction. These are, chiefly, theoretical predictions, but some
experimental findings are presented too.

Solitons are also well-known states in 1D models with nonlocal nonlinearity
(Krolikowski and Bang, 2000). Although 1D solitons are not considered in
this article in detail, it is relevant to mention the Benjamin-Ono equation
(Benjamin, 1967; Ono, 1975), which was derived as a modification of the KdV
equation for internal waves in stratified fluids, featuring nonlocal
dispersion (i.e., the nonlocality appears in the linear part of the
equation). This is an integrable equation which, similar to its KdV
counterpart, admits exact multi-soliton solutions (Ablowitz and Fokas, 1983;
Kaup and Matsumo, 1998).

The review is not designed to be a comprehensive one, as otherwise it would
grow into a full-size book. Particular topics selected for the inclusion in
the review correspond to items which are singled out in the table of
contents. Some of them represent themes and results which are well known
from previous works, therefore they are briefly outlined in the review.
Settings which were elaborated recently (especially those addressed in
Section IV, produced by a 2D system for a binary BEC whose components are
resonantly coupled by the interaction with a microwave electromagnetic
field) are presented here in a more detailed form. Some topics which are not
included in the review are mentioned in the concluding section.

\subsection{Established models: thermal and liquid-crystal (orientational)
nonlinearities in optics}

The possibility to stabilize multidimensional solitons in models with
nonlocal nonlinear terms is well known. One of the first predictions of this
stabilization mechanism for 2D solitons was published by Turitsyn (1985).
Later, this topic was elaborated in detail theoretically, see a review by
Krolikowski \textit{et al}. (2004). In optics, the nonlocal propagation is
realized in media with thermal nonlinearity, which originates form the local
variation of the refractive index due to heating the medium by the
propagating light. The corresponding model is based on the linear equation
for the paraxial propagation of optical amplitude $U\left( x,y,z\right) $
along the $z$ axis, with transverse coordinates $\left( x,y\right) $, which
is coupled to the equation for the local perturbation of the refractive
index, $n\left( x,y,z\right) $:%
\begin{equation}
iU_{z}+\frac{1}{2}\left( U_{xx}+U_{yy}\right) +nU=0,  \tag{1}
\end{equation}%
\begin{equation}
\sigma ^{-2}\left( n_{xx}+n_{yy}\right) -n=-|U|^{2}.  \tag{2}
\end{equation}%
In Eq. (2), term $-|U|^{2}$ represents the local source of heating, and $%
\sigma $ is the characteristic correlation length of the nonlocal
interaction (in the limit of $\sigma \rightarrow \infty $, the nonlocal
nonlinear system is sometimes replaced by a linear Schr\"{o}dinger equation,
with the harmonic-oscillator potential whose strength is proportional to the
norm of the wave field -- the so-called model of \textquotedblleft
accessible solitons" (Snyder and Mitchell, 1997).

The basic model for nonlinear light propagation in nematic liquid crystals
amounts to a system of equations which is similar to Eqs. (1) and (2), also
leading to the creation of stable 2D solitons (Minzoni, Smyth, and Worthy,
2007; Khoo, 2009; Assanto \textit{et al}., 2009; Peccianto and Assanto,
2012). In that case, the nonlinearity is orientational, related to rotation
of long molecules in the optical field.

The system of Eqs. (1) and (2) can be reduced to a single nonlocal NLS
equation,%
\begin{equation}
iU_{z}+\frac{1}{2}\left( U_{xx}+U_{yy}\right) +U\left( x,y\right) \int \int
G\left( \sqrt{\left( x-x^{\prime }\right) ^{2}+\left( y-y^{\prime }\right)
^{2}}\right) \left\vert U\left( x^{\prime },y^{\prime }\right) \right\vert
^{2}dx^{\prime }dy^{\prime }=0,  \tag{3}
\end{equation}%
where $G$ is the Green's function of the linear operator%
\begin{equation}
\hat{L}=-\sigma ^{2}\left( \partial _{x}^{2}+\partial _{y}^{2}\right) +1.
\tag{4}
\end{equation}%
In many works, the Green's function is approximately replaced by a Gaussian
kernel:%
\begin{equation}
G(r)=\left( \pi \sigma \right) ^{-1}\exp \left( -r^{2}/\sigma ^{2}\right) .
\tag{5}
\end{equation}%
The full system of equations (1) and (2) and the single equation (3) with
the simplified kernel (5) produce quite similar results (Wyller \textit{et al%
}., 2002).

A 3D NLS equation for the spatiotemporal propagation of light in an ionized
medium, with the cubic term which is nonlocal in the temporal coordinate,
thus representing the intra-pulse Raman shift, was introduced by Khalyapin
and Bugay (2022). The equation also includes the third-order group-velocity
dispersion, but does not include absorption of light. In the framework of
this model, stable propagation of axially symmetric \textquotedblleft light
bullets" (as spatiotemporal solitons were named by Silberberg, 1990) was
predicted, using the approximation based on the method of moments. In this
context, the full 3D NLS equation was replaced by a system of evolution
equations for seven integral moments.

In addition to straightforward stabilization of fundamental solitons, it was
found that the nonlocal model provides for the existence of stable solitons
with embedded vorticity, alias VRs (vortex rings), as well as higher-order
VRs with a multiple-ring transverse structure (Yakimenko, Zaliznyak, and
Kivshar, 2005; Briedis \textit{et al}., 2005; Skupin \textit{et al}., 2006).
The same model predicts 2D soliton complexes in the form of rotating dipoles
(Lopez-Aguayo \textit{et al}., 2006). A stability domain of vortex solitons
was also found in a model of the light propagation in liquid crystals, where
the nonlinearity is nonlocal too (Jung \textit{et al}., 2021). Therefore,
the nonlocality is a promising method for the stabilization of complex 2D
states. On the other hand, in the 3D version of the nonlocal system,
produced by adding term $U_{tt}$ to Eq. (1), where $t$ is the temporal
coordinate, only fundamental 3D solitons are stable, while VRs are not
(Mihalache \textit{et al}., 2006). In addition to vortex solitons, robust
necklace-shaped patterns in a nonlocal medium, produced by instability
development of the solitary vortices, were explored by Walasik \textit{et al}%
. (2017).

\begin{figure}[tbp]
\begin{center}
\includegraphics[width=0.60\textwidth]{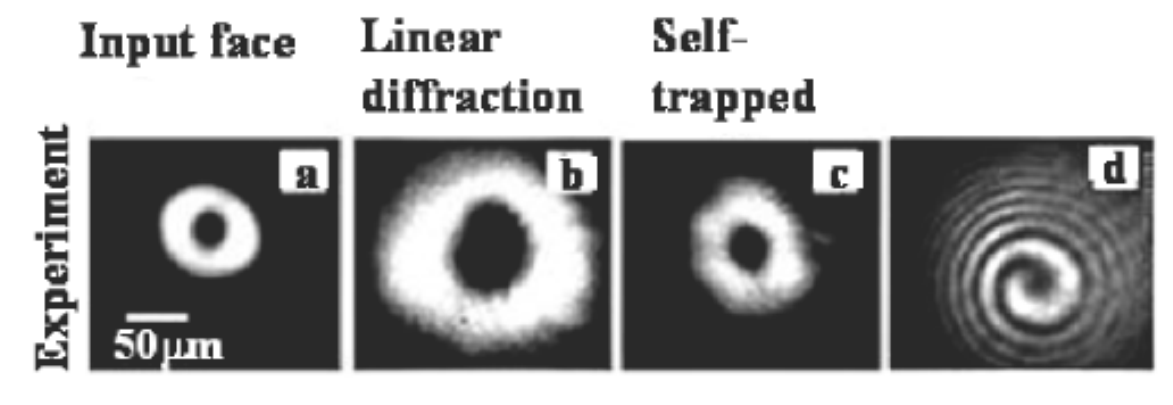}\\[0pt]
\includegraphics[width=0.60\textwidth]{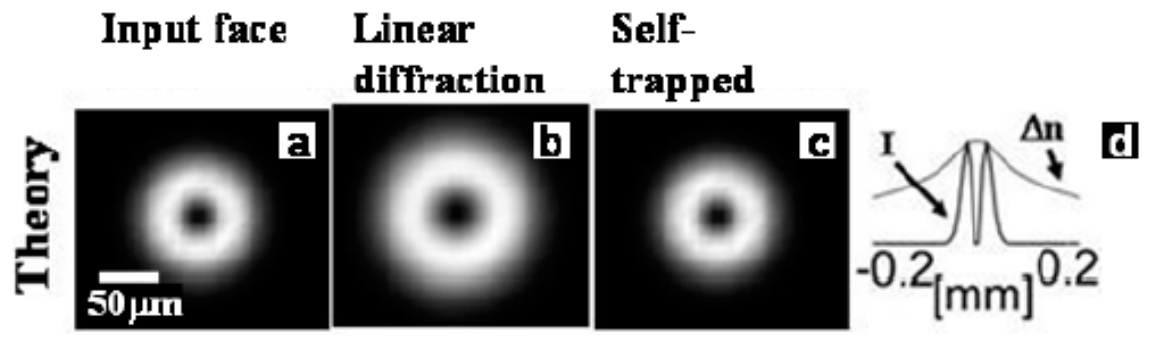}
\end{center}
\caption{Experimentally demonstrated creation of stable optical vortex
solitons (with winding number $S=1$) supported by the nonlocal nonlinearity
in the bulk waveguide made of lead glass. The top row demonstrates the
intensity distribution in the input vortex beam (a), its linear diffraction
when the power is insufficient to make the propagation nonlinear (b), and
the formation of the stable vortex soliton when the power is sufficient for
that (c). Panel (d) in the top row displays the experimentally measured
phase distribution in the vortex. Panels (a)-(c) in the bottom row show the
findings for the intensity distribution as produced, for the same setup, by
simulations of Eqs. (1) and (2). In addition to that, panel (d) in the
bottom row displays the distribution of the local intensity, $\mathrm{I}$,
and local perturbation of the refractive index, $n$ (here denoted $\mathrm{%
\Delta n}$), in the cross-section of the vortex soliton, as produced by the
numerical solution (source: Rotschield \textit{et al}., 2005).}
\label{fig-Rotschield}
\end{figure}

In the experiment, the formation of stable 2D fundamental solitons by light
beams propagating through a vapor of hot sodium atoms, which is an
effectively nonlocal optical medium, was demonstrated by Suter and Blasberg
(1993). In liquid crystals, 2D solitons, which are often called
\textquotedblleft nematicons", were created by Peccianti \textit{et al}.
(2000). Then, as illustrated by Fig. \ref{fig-Rotschield}, stable optical
VRs and elliptically deformed fundamental 2D solitons were created by
Rotschild \textit{et al}. (2005) in the lead glass, whose nonlocal
nonlinearity is adequately modelled by Eqs. (1) and (2). Stable vortex
solitons of higher orders -- in particular, with vorticity $S=4$ (Zhang
\textit{et al}., 2019) and $S=10$ (Zhang, Zhou, and Dai, 2022) -- were
experimentally demonstrated in glasses with the thermal nonlocal
nonlinearity. Stable vector (two-component) optical vortex solitons in a
liquid-crystal medium were demonstrated by Izdebskaya, Assanto, and
Krolikowski (2015).

Another experiment with the propagation of (2+1) optical beams in an
isotropic glass waveguide with the thermal nonlocal nonlinearity, and
respective simulations of Eq. (3), were performed by Zhang \textit{et al}.
(2022). They considered beams with an elliptical transverse shape. If the
elliptically deformed beam did not carry angular momentum, it performed
shape oscillations, similar to those demonstrated in simulations of the
evolution of an elliptic ring kinks in the two-dimensional SG equation
(Christiansen \textit{et al}., 1997). On the other hand, a similar input to
which angular momentum was imparted could self-trap into a stably rotating
elliptic soliton, which resembles the so-called \textquotedblleft propeller
modes" (Carmon \textit{et al}., 2001).

\subsection{A new model with \emph{linear nonlocality}: fractional
diffraction in 2D}

The introduction of fractional calculus in NLS equations has drawn much
interest since it was proposed by Laskin (2000) -- originally, in the linear
form -- as the quantum-mechanical model, derived from the Feynman-integral
formulation for particles moving by \textit{L\'{e}vy flights} (see also the
book by Laskin (2018)). Then, realization of the effective fractional
diffraction in optical cavities was proposed by Longhi (2015) and by Zhang
\textit{et al}. (2015a). Implementation of fractional linear Schr\"{o}dinger
equations in condensed-matter settings has been reported by Stickler (2013)
and by Pinsker \textit{et al}. (2015).

The nonlinearity was added to the fractional Schr\"{o}dinger equations,
starting from the work by Secchi and Squassima (2014). In terms of the
realization of the fractional diffraction in optics, the cubic self-focusing
represents the Kerr nonlinearity of the material of the waveguide. The cubic
nonlinearity also makes sense if added to the original fractional Schr\"{o}%
dinger equation in quantum mechanics. The so extended fractional model may
be considered as an effective Gross-Pitaevskii (GP) equation for the gas of
quantum particles moving by L\'{e}vy flights\textit{.} The fractional NLS
equations give rise to many theoretical results for solitons in framework of
the fractional NLS equations, see a brief review of the topic by Malomed
(2021). In particular, the 2D fractional NLS equation for amplitude $\psi
(x,y,z)$ of the optical wave propagating along the $z$ direction, under the
action of an effective transverse potential, $U(x,y)$, and the usual cubic
self-focusing, is written as%
\begin{equation}
i\frac{\partial \psi }{\partial z}=\frac{1}{2}\left( -\frac{\partial ^{2}}{%
\partial x^{2}}-\frac{\partial ^{2}}{\partial y^{2}}\right) ^{\alpha /2}\psi
-|\psi |^{2}\psi +U\left( x,y\right) \psi .  \tag{6}
\end{equation}%
The fractionality of the diffraction operator in Eq. (6) is determined by
the L\'{e}vy index (LI), $\alpha $. In usually considered models, it takes
values in interval
\begin{equation}
1\leq \alpha \leq 2,  \tag{7}
\end{equation}%
$\alpha =2$ corresponding to the usual NLS equation. Equation (6) reduces to
the 1D form by dropping coordinate $y$, which leads to equation
\begin{equation}
i\frac{\partial \psi }{\partial z}=\frac{1}{2}\left( -\frac{\partial ^{2}}{%
\partial x^{2}}\right) ^{\alpha /2}\psi -|\psi |^{2}\psi +U\left( x\right)
\psi .  \tag{8}
\end{equation}

The affinity of Eqs. (6) and (8) to nonlocal models, although with \emph{%
linear} nonlocality, is demonstrated by the definition of the fractional
derivative in Eq. (8), and the fractional-diffraction operator in Eq. (6):%
\begin{equation}
\left( -\frac{\partial ^{2}}{\partial x^{2}}\right) ^{\alpha /2}\psi (x)=%
\frac{1}{2\pi }\int_{-\infty }^{+\infty }dp|p|^{\alpha }\int_{-\infty
}^{+\infty }d\xi e^{ip(x-\xi )}\psi (\xi ),  \tag{9}
\end{equation}%
\begin{equation}
\left( -\frac{\partial ^{2}}{\partial x^{2}}-\frac{\partial ^{2}}{\partial
y^{2}}\right) ^{\alpha /2}\psi (x,y)=\frac{1}{(2\pi )^{2}}\int \int
dpdq\left( p^{2}+q^{2}\right) ^{\alpha /2}\int \int d\xi d\eta e^{i\left[
p(x-\xi )+iq(y-\eta )\right] }\psi (\xi ,\eta ).  \tag{10}
\end{equation}%
These integral expressions are produced, essentially, as juxtapositions of
the direct and inverse Fourier transform for field $\psi $. In fact, there
are many different definitions of fractional derivatives; the one adopted in
Eq. (9), which is relevant to the above-mentioned physical realizations in
quantum mechanics and optics, is often called the \textit{Riesz derivative}
(Agrawal, 2007; Cai and Li, 2019). It is relevant to mention that
alternative definitions of the fractional derivatives, such as Caputo and
Riemann-Liouville ones, may also produce solitons in the framework of the
fractional NLS equation (Kwasnicki, 20017; Navickas \textit{et al}., 2017),
although the use of those definitions in the context of optics models is
less straightforward.

An essential problem posed by the fractional NLS equations (6) and (8),
which include the cubic self-focusing term, is the possibility of the onset
of collapse in it. Well-known criteria for the occurrence of the
supercritical and critical collapse for the usual NLS equations (Berg\'{e},
1998; Sulem and Sulem, 1999; Fibich, 2015) can be easily generalized for the
fractional NLS equations (6) and (8) with the cubic self-focusing. The
derivation is based on the analysis of scaling of the fractional-diffraction
and self-focusing terms, $E_{\mathrm{diffr}}>0$ and $E_{\mathrm{focus}}<0$
in the energy Hamiltonian of the equation, assuming spontaneous
self-compression of the wave-function configuration towards the limit of the
zero spatial scale, $L\rightarrow 0$, under the condition of the
conservation of the integral norm, $\mathcal{N}=\int \left\vert \psi (%
\mathbf{r})\right\vert ^{2}d\mathbf{r}$. The latter condition implies that
the squared amplitude of the wave function scales as $A^{2}\sim \mathcal{N}%
/L^{D}$, where $D=1$ and $2$ for Eqs. (6) and (8), respectively. With regard
to this result, the conclusions for the scaling are $E_{\mathrm{diffr}}\sim
\mathcal{N}/L^{\alpha }$ and $E_{\mathrm{focus}}\sim -\mathcal{N}^{2}/L^{D}$%
. The collapse cannot develop if $E_{\mathrm{diffr}}$ grows at $L\rightarrow
0$ faster than $\left\vert E_{\mathrm{focus}}\right\vert $. Thus, the
conclusion is that the critical and supercritical collapse takes place,
severally, at LI values%
\begin{equation}
\alpha _{\mathrm{crit}}=D,\alpha _{\mathrm{supercrit}}<D.  \tag{11}
\end{equation}%
In other words, in the case of the 1D fractional equation (8), the critical
collapse takes place at $\alpha =1$, and does not occur at $\alpha >1$. In
the 2D fractional model based on Eq. (6), the collapse takes place in the
entire interval (7): critical at $\alpha =2$ (which is tantamount to the
usual 2D NLS equation with the cubic self-focusing), and supercritical at $%
\alpha <2$.

The occurrence of the collapse makes it difficult to obtain stable solitons
as solutions of Eq. (6). Nevertheless, stable 2D solitons, including ones
with embedded vorticity, were predicted, adding the self-defocusing quintic
term to the model (Li, Malomed, and Mihalache, 2020(a,b)), or replacing the
local cubic self-focusing by its nonlocal counterpart, the same as in Eqs.
(3) and (5). A trapping harmonic-oscillator potential in Eq. (6), $%
U(x,y)=\left( \Omega ^{2}/2\right) \left( x^{2}+y^{2}\right) $, may also
provide stabilization of 2D solitons at $\alpha <2$ (Malomed, 2021). The
latter result is similar to the well-known one for the usual two-dimensional
NLS/GP equation, which corresponds to $\alpha =2$ in Eq. (6). Further, a
similar fractional model with a double-well potential gives rise to 2D
solitons with spontaneously broken symmetry (Li, Li, and Dai, 2021).

\section{Anisotropic quasi-2D solitons built by dipole-dipole interactions
(DDIs) in BEC of magnetic atoms}

An important example of nonlocal nonlinearity is provided by the DDI of
magnetic atoms in BEC composed of such atoms (Lahaye \textit{et al}., 2009).
The corresponding GP equation, for the mean-field wave function $\psi $,
includes both the local nonlinearity, with strength $g$, that represents
contact collisions between atoms, and the nonlocal DDI, with strength $g_{%
\mathrm{DDI}}>0$. In the scaled form, the GP equation is%
\begin{equation}
i\frac{\partial \psi }{\partial t}=-\frac{1}{2}\nabla ^{2}\psi +g|\psi
|^{2}\psi +g_{\mathrm{DDI}}\psi (\mathbf{r})\int \left( 1-\frac{3\left(
z-z^{\prime }\right) ^{2}}{\left\vert \mathbf{r-r}^{\prime }\right\vert ^{2}}%
\right) \left\vert \psi \left( \mathbf{r}^{\prime }\right) \right\vert \frac{%
d\mathbf{r}^{\prime }}{\left\vert \mathbf{r-r}^{\prime }\right\vert ^{3}}+U(%
\mathbf{r})\psi ,  \tag{12}
\end{equation}%
where $z$ is the direction in which atomic magnetic moments are polarized by
an external magnetic field, $\int d\mathbf{r}^{\prime }$ stands for the 3D
integration, and $U(\mathbf{r})$ is an external potential acting on the
condensate. In terms of physical units, $g=4\pi a_{s}N/r_{0}$ and $g_{%
\mathrm{DDI}}=\mu _{0}\mu ^{2}mN/\left( 4\pi \hbar ^{2}r_{0}\right) $, where
$a_{s}$ is the scattering length of the atonic collisions, $N$ the number of
atoms in the condensate, $r_{0}$ a characteristic spatial scale, $\mu _{0}$
the vacuum permeability, $m$ the atomic mass, and $\mu $ the atomic magnetic
moment.

Stable 3D solitons in the free space ($U=0$) cannot be supported by the DDI.
A possibility is to create quasi-2D (pancake-shaped) solitons, applying the
confining potential acting in a particular direction. The simplest
configuration of that type is an axially symmetric one, with the potential
applied along the same axis, $z$, along which the polarizing field is
directed. However, self-trapping of the wave function in this configuration
is impossible because the interaction between parallel magnetic dipoles is
repulsive. On the other hand, it was predicted by Giovanazzi, G\"{o}rlitz,
and Pfau (2003) that the sign of the DDI may be effectively reversed,
replacing $g_{\mathrm{DDI}}\rightarrow -g_{\mathrm{DDI}}$ in Eq. (12), by
means of an additional ac magnetic field which drives fast rotation of the
dipoles (a similar prediction for polar molecules carrying electric dipole
moments was made by Micheli \textit{et al}. (2007)). In the framework of
this setup, stable solitons were predicted by Pedri and Santos (2005), and
stable vortex solitons with winding number $S=1$ were predicted by
Tikhonenkov, Malomed, and Vardi (2008b).

\begin{figure}[tbp]
\centering{\includegraphics[scale=0.5]{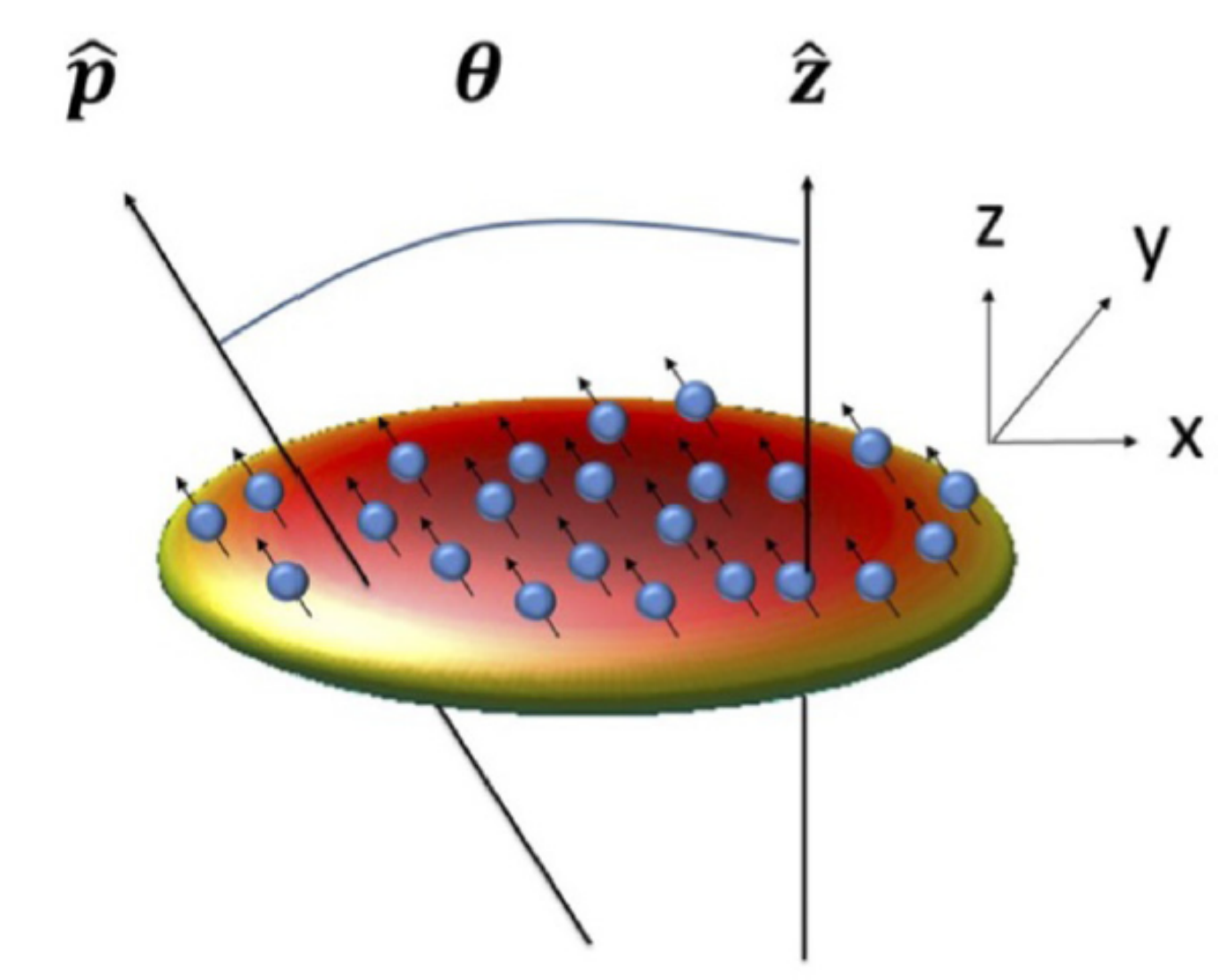}}
\caption{A scheme for the creation of quasi-2D (\textquotedblleft
pancake-shaped") solitons supported by the DDI of magnetic atoms confined in
the vertical direction by the trapping harmonic-oscillator potential. The
angle between the fixed orientation of atomic magnetic dipoles, $\mathbf{%
\hat{p}}$, and the confinement direction is $\protect\theta $ (source: Chen
\textit{et al}., 2017).}
\label{fig13.1=fig159}
\end{figure}

A more challenging objective is to search for stable quasi-2D solitons,
supported by the DDI proper, with the magnetic moments polarized
perpendicular to the confinement direction, or at an angle $\theta $ to it,
as shown schematically in Fig. \ref{fig13.1=fig159} (the notation for the
Cartesian coordinates in this figure is different from that in Eq. (12)).
An approach to the solution of this problem was elaborated by Tikhonenkov,
Malomed, and Vardi (2008a), assuming that the confinement was imposed in the
direction of $y$ by the harmonic-oscillator potential
\begin{equation}
U(\mathbf{r})=\left( \Omega ^{2}/2\right) y^{2}  \tag{13}
\end{equation}%
in Eq. (12) (i.e., indeed, the confinement direction is perpendicular to the
polarizing magnetic field). The consideration started with the VA, based on
an anisotropic Gaussian ansatz for the 3D wave function,%
\begin{equation}
\psi _{\mathrm{aniso}}=\pi ^{-3/4}\left( \alpha \beta \gamma \right)
^{1/4}\exp \left[ i\mu t-\frac{1}{2}\left( \alpha x^{2}+\beta y^{2}+\gamma
z^{2}\right) \right] ,  \tag{14}
\end{equation}%
supplemented by the normalization condition,%
\begin{equation}
\int \left\vert \psi (\mathbf{r})\right\vert d\mathbf{r}=1.  \tag{15}
\end{equation}

Ansatz (14) was inserted in the energy (Hamiltonian) corresponding to Eq.
(12), with the confining potential (13) (i.e., the atomic magnetic dipoles
are polarized along axis $z$, and the atoms are located close to the $\left(
z,x\right) $ plane:
\begin{equation}
E=E_{\mathrm{local}}+E_{\mathrm{nonlocal}},  \tag{16}
\end{equation}%
\begin{equation}
E_{\mathrm{local}}=\frac{1}{2}\int \left( |\nabla \psi (\mathbf{r})|^{2}+%
\frac{\Omega ^{2}}{2}y^{2}|\psi (\mathbf{r})|^{2}+g|\psi (\mathbf{r}%
)|^{4}\right) d\mathbf{r,}  \tag{17}
\end{equation}%
\begin{equation}
E_{\mathrm{nonlocal}}=\frac{1}{2}g_{\mathrm{DDI}}\int \int \left[ 1-\frac{%
3\left( z-z^{\prime }\right) ^{2}}{\left\vert \mathbf{r}-\mathbf{r^{\prime }}%
\right\vert ^{2}}\right] |\psi (\mathbf{r^{\prime }})|^{2}|\psi (\mathbf{r}%
)|^{2}\frac{d\mathbf{r}d\mathbf{r^{\prime }}}{\left\vert \mathbf{r}-\mathbf{%
r^{\prime }}\right\vert ^{3}}.  \tag{18}
\end{equation}%
The substitution of ansatz (14) and calculation of the integrals yields%
\begin{equation}
E_{\mathrm{VA}}=\frac{1}{4}(\alpha +\beta +\gamma )+\frac{1}{4\beta }+\sqrt{%
\frac{\alpha \beta \gamma }{2\pi }}\left[ \frac{g}{4\pi }+\frac{g_{\mathrm{%
DDI}}}{3}h(\kappa _{x},\kappa _{y})\right] ,  \tag{19}
\end{equation}%
where%
\begin{equation}
\kappa _{x}\equiv \sqrt{\gamma /\alpha },\kappa _{y}\equiv \sqrt{\gamma
/\beta },  \tag{20}
\end{equation}%
\begin{equation}
h(\kappa _{x},\kappa _{y})\equiv \int_{0}^{1}\frac{3\kappa _{x}\kappa
_{y}x^{2}dx}{\sqrt{1+(\kappa _{x}^{2}-1)x^{2}}\sqrt{1+(\kappa
_{y}^{2}-1)x^{2}}}-1.  \tag{21}
\end{equation}%
Then, values of the variational parameters $\alpha $, $\beta $, $\gamma $ in
ansatz (14) are predicted by the energy-minimization condition,%
\begin{equation}
\partial E_{\mathrm{VA}}/\partial \left( \alpha ,\beta ,\gamma \right) =0.
\tag{22}
\end{equation}

Detailed analysis of Eq. (22) has demonstrated that, under condition%
\begin{equation}
g/g_{\mathrm{DDI}}<4\pi /3\approx 4.19,  \tag{23}
\end{equation}%
it yields a minimum of energy (19), which may predict the existence of
stable solitons as solutions to Eq. (12). The meaning of constraint (23) is
that, naturally, stable solitons cannot exist if the DDI is not strong
enough.

Direct numerical solution of Eq. (12) has produced stable solitons with
shapes close to those predicted by the VA, see an example in Fig. \ref%
{fig13.2=fig160}. The strongly anisotropic shape of the soliton is a natural
manifestation of the anisotropic form of the DDI in Eq. (12).

\begin{figure}[tbp]
\centering{\includegraphics[scale=0.6]{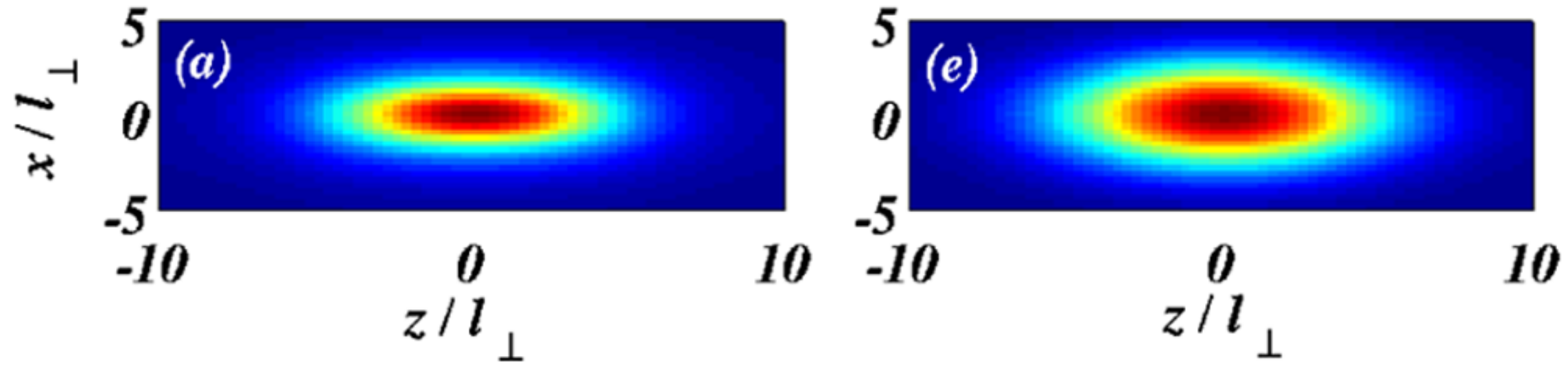}}
\caption{The right panel: the density distribution in a stable quasi-2D
(pancake-shaped) soliton solution of Eq. (12) in the mid plane, $y=0$, for
parameters $g=10$ and $g/g_{\mathrm{DDI}}=0.911$ (which satisfies condition
(23)). The right panel: the same, as predicted by the VA solution based on
ansatz (14) (source: Tikhonenkov, Malomed, and Vardi, 2008a).}
\label{fig13.2=fig160}
\end{figure}

\begin{figure}[tbp]
\centering{\includegraphics[scale=0.9]{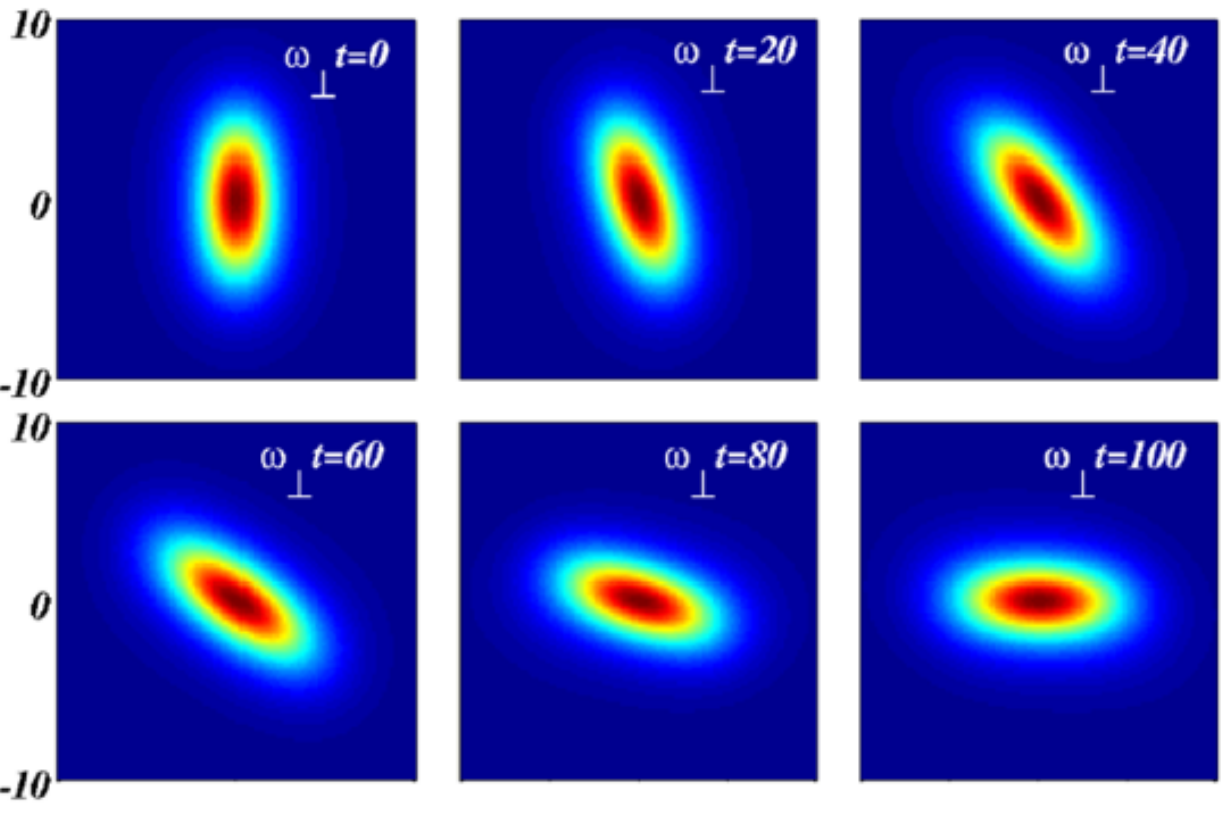}}
\caption{Stable counter-clockwise rotation of the pancake-shaped soliton
(the same whose static shape is shown in Fig. \protect\ref{fig13.2=fig160}),
following the rotation of the coordinates as per Eq. (24) with angular
velocity $\protect\omega =\protect\pi /200$. Shown are snapshots of the
density distribution in the mid plane, produced by simulations of Eq. (12)
at moments of times indicated in the panels (source: Tikhonenkov, Malomed,
and Vardi, 2008a).}
\label{fig13.3=fig161}
\end{figure}

Because the polarization of the atomic magnetic moments in the $\left(
x,z\right) $ plane is determined by the orientation of the external magnetic
field, it is also interesting to explore a dynamical state in which the
field slowly rotates in this plane (perpendicular to the confinement
direction, $y$). The dynamics can be simulated using the GP equation (12),
in which the DDI term is written in the rotating coordinates,%
\begin{equation}
x^{\prime }=x\cos \left( \omega t\right) +y\sin \left( \omega t\right)
,y^{\prime }=y\cos \left( \omega t\right) -x\sin \left( \omega t\right) .
\tag{24}
\end{equation}%
The simulations have demonstrated that the stable pancake-shaped soliton is
able to adiabatically follow the rotation of the polarizing field, provided
that the angular velocity $\omega $ in Eq. (24) is small enough, as shown in
Fig. \ref{fig13.3=fig161}. Faster rotation leads to deformation of the
soliton.

An experimentally realistic scenario for the creation of the stable
pancake-shaped solitons was elaborated by K\"{o}berle \textit{et al}.
(2012). The scenario includes temporal variation of the scattering length
and strength of the trapping potential, with the purpose to help building
the soliton from an experimentally available input. Also included were
additional physically significant factors, such as were three-body losses
and random fluctuations of the scattering length.

A similar, but more general, quasi-2D setting, under the action of the
confining potential (13) with the tilt angle $\theta <\pi /2$ in Fig. \ref%
{fig13.1=fig159}, was considered by Chen \textit{et al}. (2017). Using a
combination of VA and numerical solution of the accordingly modified Eq.
(12), it was found that stable pancake-shaped solitons with the anisotropic
shape persist in an interval of the tilt angles \
\begin{equation}
\theta _{\max }<\theta \leq \pi /2.  \tag{25}
\end{equation}%
The limit value $\theta _{\max }$ in Eq. (25) depends on parameters, always
staying close to the so-called \textit{magic angle}, \
\begin{equation}
\theta _{\mathrm{magic}}=\arccos \left( 1/\sqrt{3}\right) \approx 54.74^{%
\mathrm{o}}.  \tag{26}
\end{equation}%
The meaning of this angle is that the potential of the DDI for two parallel
point-like dipoles, with angle $\theta $ between the line of length $R$
connecting them and the common direction of the magnetic moments, is
proportional to%
\begin{equation}
\mathrm{Potential(DDI)}\sim R^{-3}\left( 1-3\cos ^{2}\theta \right) .
\tag{27}
\end{equation}%
Thus, potential (27) vanishes at $\theta =\theta _{\mathrm{magic}}$.

Equation (12) with potential (13) keeps the Galilean invariance in the $%
\left( x,z\right) $ plane, which suggests to set the anisotropic solitons in
motion in this plane, and simulate collisions between them (Eichler \textit{%
et al}., 2011). The simulations demonstrate that the collisions may be
deeply inelastic, leading to merger of the colliding solitons into a single
quasi-soliton state (Young-S. and Adhikari, 2022).

Recently, a 3D spatiotemporal model similar to Eq. (12) was derived by Zhao
\textit{et al}. (2022) for a two-component model of a gas of Rydberg atoms
with long-range interactions in an optical medium with
electromagnetically-induced transparency. The model produces stable 3D
fundamental solitons, as well as stable solitons with embedded vorticity. A
protocol for storage and retrieval of the 3D solitons in this system was
elaborated too.

\section{Giant vortex rings (VRs) in microwave-coupled binary BEC}

Photonic tools, such as optical-lattice and trapping potentials, are broadly
used in the experimental and theoretical work with single- and
multi-component BECs. In particular, microwave (MW) fields are used to
resonantly couple different atomic states which form two-component
condensates (Ballagh, Burnett, and Scott, 1997; Bookjans, Vinit, and Raman,
2011). In many cases, the feedback of the BEC on the MW fields is ignored.
Nevertheless, the feedback produced by relatively dense condensates may
induce field-mediated long-range interaction in BEC, which is often called
the \textit{local-field effect} (LFE) and gives rise to significant
phenomena. In particular, the LFE acting on the electric component of the
field explains asymmetric matter-wave diffraction (Li \textit{et al}. 2008;
Zhu \textit{et al}., 2011) and predicts polaritonic solitons in soft optical
lattices (Dong \textit{et al}., 2013). Further, the resonant coupling of the
magnetic components of the MW field to the condensate of two-level atoms
opens the way to the creation of hybrid microwave-matter-wave solitons (Qin,
Dong, and Malomed, 2015). Actually, the MW-mediated long-range interaction
may cover the whole condensate, in contrast with fast-decaying nonlocal
interactions in optics and dipolar BEC (cf. Eqs. (3), (5) and (12)).

In the 2D configuration, a hybrid system including a pseudo-spinor BEC
matter-wave function, whose two components are coupled by the MW field, as
shown schematically in Fig. \ref{fig13.4=fig162}, was introduced by Qin,
Dong, and Malomed (2016). As shown below, stable solitons in the form of
VRs, with arbitrarily large values of winding number $S$, readily self-trap
in this 2D setting. The conclusion remains valid if the system includes the
local repulsive or attractive interaction. In particular, the domain in
which the VRs remain stable against the critical collapse, driven by the
local attraction between the components, \emph{expands} with the increase of
$S$, persisting for \emph{arbitrarily high} values of $S$. This conclusion
is remarkable, as, in other systems which admit stable VRs with $S>1$, their
stability domain shrinks with the increase of $S$. In this sense, the VRs
predicted in the 2D hybrid matter-wave-microwave system may be considered as
stable \emph{giant vortices}, because large values of $S$ naturally imply a
large radius of the ring.
\begin{figure}[tbp]
\centering\includegraphics[scale=0.75]{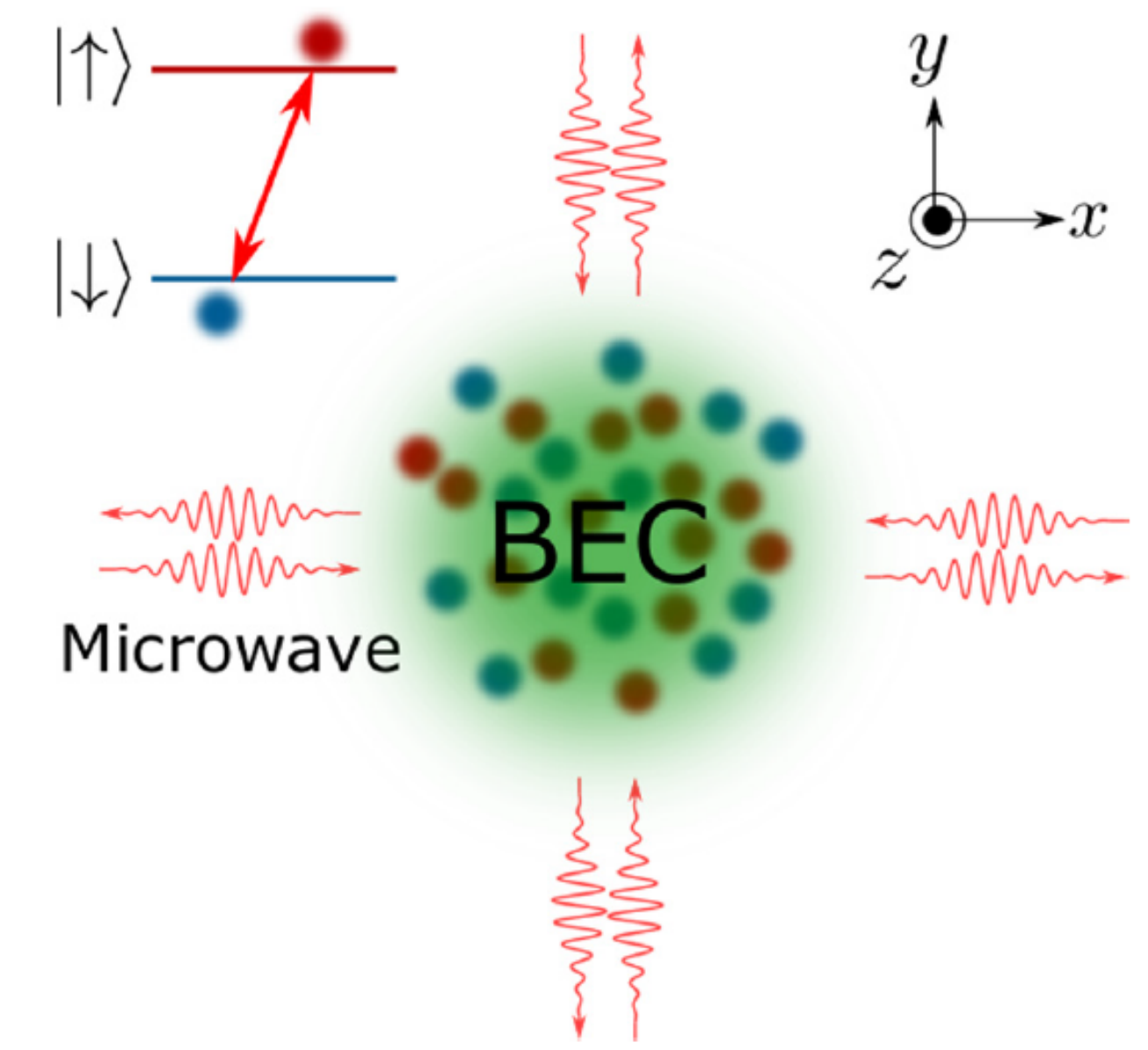}
\caption{The scheme of the 2D hybrid system which builds giant VRs. The
pseudo-spinor (two-component) wave function represents two atomic states
coupled by the MW (microwave) field, as shown in the figure. The MW field is
polarized in the direction perpendicular to the system's plane (source: Qin,
Dong, and Malomed, 2016). }
\label{fig13.4=fig162}
\end{figure}

\subsection{The model}

Following Fig. \ref{fig13.4=fig162}, a nearly-2D binary BEC, composed of two
different atomic states, is described by the pseudo-spinor wave function, $%
\left\vert \Phi \right\rangle =\left( \phi _{\downarrow },\phi _{\uparrow
}\right) ^{T}$, with each component emulating \textquotedblleft spin-up" and
\textquotedblleft spin-down" states of the usual spinor. In the scaled
notation (setting the Planck's constant, atomic mass, vacuum magnetic
permeability, and the absolute value of the magnetic moment to be $1$), the
corresponding atomic Hamiltonian is
\begin{equation}
\mathcal{H}=\hat{\mathbf{p}}^{2}/2+\eta \sigma _{3}-\mathbf{m}\cdot \mathbf{%
B,}  \tag{28}
\end{equation}%
where $\hat{\mathbf{p}}$ and $\mathbf{m}$\ are the 2D momentum and magnetic
moment, $2\eta $\ is detuning of the MW from the transition between the
atomic states $\left\vert \uparrow \right\rangle $\ and $\left\vert
\downarrow \right\rangle $, $\sigma _{3}$\ is the Pauli matrix, and
\begin{equation}
\mathbf{B}=\mathbf{H}+\mathbf{M}  \tag{29}
\end{equation}%
is the magnetic induction, with magnetic field $\mathbf{H}$\ and
magnetization $\mathbf{M}=\left\langle \Phi \right\vert \mathbf{m}\left\vert
\Phi \right\rangle $. Assuming that the atomic magnetic moments are
polarized along the field, the field and magnetization may be taken in the
scalar form. Then, in the rotating-wave approximation the components of the
wave function obey the following system of coupled GP equations:%
\begin{equation}
i\frac{\partial \phi _{\downarrow }}{\partial t}=\left( -\frac{1}{2}\nabla
^{2}+\eta \mathbf{-}\beta \left\vert \phi _{\uparrow }\right\vert
^{2}\right) \phi _{\downarrow }-\gamma H^{\ast }\phi _{\uparrow },  \tag{30}
\end{equation}%
\begin{equation}
i\frac{\partial \phi _{\uparrow }}{\partial t}=\left( -\frac{1}{2}\nabla
^{2}-\eta \mathbf{-}\beta \left\vert \phi _{\downarrow }\right\vert
^{2}\right) \phi _{\uparrow }-\gamma H\phi _{\downarrow }.  \tag{31}
\end{equation}%
Here $\ast $ stands for the complex conjugate, $\gamma $ is the strength of
the MW-atom coupling, and the strength of the cross-interaction of the two
components is determined by the scalar product of the matrix elements of the
magnetic moment: $\beta =\mathbf{m}_{\uparrow \downarrow }\cdot \mathbf{m}%
_{\downarrow \uparrow }$.

The magnetic field is determined by the inhomogeneous Helmholtz equation. In
the present notation, it is
\begin{equation}
\nabla ^{2}H+k^{2}H=-\phi _{\downarrow }^{\ast }\phi _{\uparrow },  \tag{32}
\end{equation}%
\ where $k$\ is the MW wavenumber. As the respective wavelength of the MW
field, $\lambda =2\pi /k$, is always much greater than an experimentally
relevant size of the BEC, the second term in Eq. (32) may be omitted in
comparison with the first term, reducing Eq. (32) to the Poisson equation:
\begin{equation}
\nabla ^{2}H=-\phi _{\downarrow }^{\ast }\phi _{\uparrow }.  \tag{33}
\end{equation}%
Because the medium's magnetization, which is the source of the magnetic
field, is concentrated in the nearly-2D \textquotedblleft pancake", the
Poisson equation may be treated as a two-dimensional one. Then, using the
Green's function of the 2D Poisson equation, the magnetic field is produced
by Eq. (33) as\textbf{\ }%
\begin{equation}
H(\mathbf{r})=-\frac{1}{2\pi }\int \!\!\ln \left( \left\vert \mathbf{r}-%
\mathbf{r^{\prime }}\right\vert \right) \phi _{\downarrow }^{\ast }\left(
\mathbf{r^{\prime }}\right) \phi _{\uparrow }\left( \mathbf{r^{\prime }}%
\right) d\mathbf{r}^{\prime }\mathbf{,}  \tag{34}
\end{equation}%
\textbf{\ }where $\mathbf{r}$\ is the set of the coordinates in the 2D
plane. The substitution of expression (34) in GP equations (30) and (31)
casts them in the form of coupled NLS equations with the nonlocal
interaction, cf. Eqs. (3) and (12), acting along with the local
cross-interaction with strength $\beta $:\textbf{\ }%
\begin{equation}
i\frac{\partial \phi _{\downarrow }}{\partial \tau }=\left( -\frac{1}{2}%
\nabla ^{2}+\eta -\beta \left\vert \phi _{\uparrow }\right\vert ^{2}\right)
\phi _{\downarrow }+\frac{\gamma \phi _{\uparrow }}{2\pi }\int \!\!\ln
\left( \left\vert \mathbf{r}-\mathbf{r^{\prime }}\right\vert \right) \phi
_{\downarrow }\left( \mathbf{r^{\prime }}\right) \phi _{\uparrow }^{\ast
}\left( \mathbf{r^{\prime }}\right) \!d\mathbf{r^{\prime }},  \tag{35}
\end{equation}%
\begin{equation}
i\frac{\partial \phi _{\uparrow }}{\partial \tau }=\left( -\frac{1}{2}\nabla
^{2}-\eta -\beta \left\vert \phi _{\downarrow }\right\vert ^{2}\right) \phi
_{\uparrow }+\frac{\gamma \phi _{\downarrow }}{2\pi }\int \!\!\ln \left(
\left\vert \mathbf{r}-\mathbf{r^{\prime }}\right\vert \right) \phi
_{\downarrow }^{\ast }\left( \mathbf{r^{\prime }}\right) \phi _{\uparrow
}\left( \mathbf{r^{\prime }}\right) \!d\mathbf{r^{\prime }},  \tag{36}
\end{equation}%
Equations (35) and (36) are supplemented by the normalization condition,
\begin{equation}
\int \left( \left\vert \phi _{\uparrow }\right\vert ^{2}+\left\vert \phi
_{\downarrow }\right\vert ^{2}\right) d\mathbf{r}=1.  \tag{37}
\end{equation}

If collisions between atoms belonging to the two components are considered
(with the corresponding strength of the contact interaction tunable, in the
experiment, by means of the Feshbach resonance), the additional
cross-interaction terms can be absorbed into rescaled coefficient $\beta $
in Eqs. (35) and (36). Collisions may also give rise to self-interaction
terms, $-\tilde{\beta}\left\vert \phi _{\downarrow }\right\vert ^{2}$ and $-%
\tilde{\beta}\left\vert \phi _{\uparrow }\right\vert ^{2}$, in the
parentheses of Eqs. (35) and (36), respectively. Note also that the same
equations (35) and (36) (without the self-interaction terms) apply to a
different physical setting, \textit{viz}., a degenerate Fermi gas with spin $%
1/2$, in which $\phi _{\downarrow }$ and $\phi _{\uparrow }$ represent two
spin components, coupled by the MW magnetic field (Radzihovsky and Sheehy,
2010; Qin, Dong, and Malomed, 2015).

The following analysis chiefly addresses the symmetric (zero-detuning)
system, which corresponds to $\eta =0$ in Eqs. (35) and (36). Then, these
equations coalesce into a single one for $\phi _{\downarrow }=\phi
_{\uparrow }\equiv \phi $,%
\begin{equation}
i\frac{\partial \phi }{\partial t}=\left[ -\frac{1}{2}\nabla ^{2}-\beta
\left\vert \phi \right\vert ^{2}+\frac{\gamma }{2\pi }\int \!\!\ln \left(
\left\vert \mathbf{r}-\mathbf{r^{\prime }}\right\vert \right) \left\vert
\phi \left( \mathbf{r^{\prime }}\right) \right\vert ^{2}\!d\mathbf{r^{\prime
}}\right] \phi ,  \tag{38}
\end{equation}%
and normalization (37) reduces to
\begin{equation}
\int \left\vert \phi (\mathbf{r})\right\vert ^{2}d\mathbf{r}=1/2.  \tag{39}
\end{equation}%
In this case, the above-mentioned self-interaction coefficient, $\tilde{\beta%
}$, may be absorbed into $\beta $.

The remaining scaling invariance of Eq. (38) makes it possible to finally
fix $\gamma =\pi $. In physical units, assuming the transverse-confinement
size $l_{\perp }\sim 1~\mathrm{\mu }$m and MW wavelength $\sim 1$\ mm,
typical solutions for VR solitons presented below correspond to
\textquotedblleft heavy" BECs with the number of atoms $N\sim 10^{8}$, which
are available in the experiment (Comparat \textit{et al}., 2006; van der
Stam \textit{et al}., 2007), a typical VR radius being $\sim 10$\ $\mathrm{%
\mu }$m.

\subsection{Results}

In polar coordinates $\left( r,\theta \right) $, stationary solutions to Eq.
(38) with chemical potential $\mu $ and integer vorticity $S$ are looked for
as
\begin{equation}
\phi =e^{-i\mu \tau -iS\theta }\Phi _{S}\left( r\right) \mathbf{,}  \tag{40}
\end{equation}%
\textbf{\ }where $\Phi _{S}(r)$ is a real radial wave function, which
satisfies the following equation, obtained by the substitution of ansatz
(40) in Eq. (38):%
\begin{equation}
\left[ \mu +\frac{1}{2}\frac{d^{2}}{dr^{2}}-\frac{S^{2}}{r^{2}}+\beta \Phi
_{S}^{2}\right] \Phi _{S}=\gamma \Phi _{S}(r)\int_{0}^{\infty }\ln \left(
\frac{1}{2}(r+r^{\prime }+\frac{1}{2}\left\vert r-r^{\prime }\right\vert
\right) \Phi _{S}^{2}(r^{\prime })r^{\prime }dr^{\prime }.  \tag{41}
\end{equation}%
The form of the nonlocal term in Eq. (41) was derived by explicitly
performing the angular integration in the last term of Eq. (38). The
corresponding magnetic field $H(r)$ is then produced by performing the
integration with respect to the angular coordinate in Eq. (34):%
\begin{equation}
H(r)=-\left[ \int_{0}^{r}\left( \ln r^{\prime }\right) +\left( \ln r\right)
\int_{r}^{\infty }\right] \Phi _{S}^{2}(r^{\prime })r^{\prime }dr^{\prime }.
\tag{34'}
\end{equation}

\begin{figure}[tbp]
\centering\includegraphics[scale=0.6]{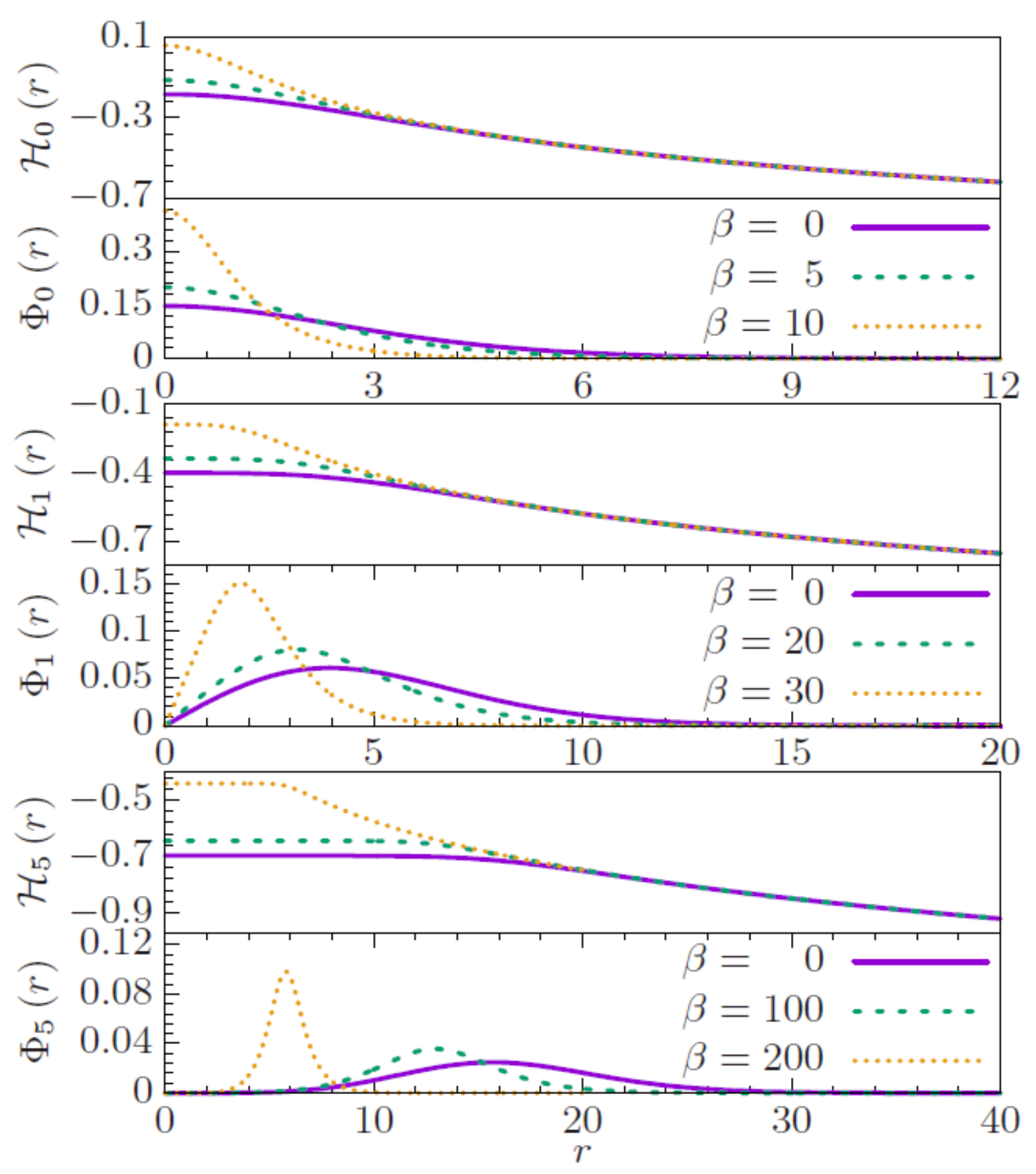}
\caption{Numerically found radial wave function $\Phi _{S}(r)$ (defined as
per Eq. (40)) and the corresponding magnetic field, $H_{S}(r)$, calculated
as per Eq. (34). Top panels: the fundamental solitons ($S=0$); middle
panels: stable VRs with $S=1$; bottom panels: stable higher-order VRs ($S=5$%
). All the solutions pertain to indicated values of strength $\protect\beta $
of the inter-component attraction, and $\protect\gamma =2\protect\pi $
(source: Qin, Dong, and Malomed, 2016).}
\label{fig13.5=fig163}
\end{figure}

Characteristic examples of solutions for $\Phi _{S}\left( r\right) $,
produced by the imaginary-time simulations of Eq. (38), along with the
corresponding profiles of $H(r)$, are displayed in Fig. \ref{fig13.5=fig163}%
, for different values of $S$\ and $\beta \geq 0$. Numerical results
demonstrate that the fundamental solitons, which correspond to $S=0$, and
VRs with $S\geq 1$ are destroyed by the collapse at $\beta >\beta _{\max
}(S) $, see Table I. This critical value of the coefficient of the
inter-component attraction can be found by considering the energy
corresponding to Eq. (38),%
\begin{equation}
E=2\pi \int_{0}^{\infty }rdr\left[ \left( \frac{d\Phi _{S}^{\prime }}{dr}%
\right) ^{2}+\frac{S^{2}}{r^{2}}\Phi _{S}^{2}-\beta \Phi _{S}^{4}\right] +%
\frac{\gamma }{2\pi }\int \int d\mathbf{r}_{1}d\mathbf{r}_{2}\ln \left(
\left\vert \mathbf{r}_{1}-\mathbf{r}_{2}\right\vert \right) \Phi _{S}^{2}(%
\mathbf{r}_{1})\Phi _{S}^{2}(\mathbf{r}_{2}).  \tag{42}
\end{equation}

The numerical findings demonstrate that, for $S\geq 2$ and $\beta $ large
enough, the vortex soliton is shaped as a narrow ring, see Fig. \ref%
{fig13.5=fig163}. It may be approximated by the usual quasi-1D soliton shape
in the radial direction (cf. the approximation which was used, in the 2D NLS
equation with the CQ nonlinearity, by Caplan \textit{et al}. (2012)):
\begin{equation}
\Phi _{S}(r)=\frac{\sqrt{\beta }}{8\pi R}\mathrm{sech}\left( \frac{\beta }{%
8\pi }\frac{r-R}{R}\right) ,  \tag{43}
\end{equation}%
where $R$ is the VR's radius, and normalization (39) is taken into regard.
Then, the substitution of approximation (43) in Eq. (42) yields%
\begin{equation}
E(R)=\left[ S^{2}-\frac{\beta ^{2}}{3\left( 8\pi \right) ^{2}}\right] \frac{1%
}{2R^{2}}+\frac{\gamma }{8\pi }\ln R.  \tag{44}
\end{equation}%
Radius $R$ of the soliton's ring is selected as a value corresponding to the
energy minimum, $dE/dR=0$, i.e.,
\begin{equation}
R_{\min }^{2}=\frac{8\pi }{\gamma }\left[ S^{2}-\frac{1}{3}\left( \frac{%
\beta }{8\pi }\right) ^{2}\right] .  \tag{45}
\end{equation}%
In comparison with numerical results, Eq. (45) provides a reasonable
approximation for the radius of narrow VRs. Then, the above-mentioned
critical value $\beta _{\max }$ is analytically (\textquotedblleft $\mathrm{%
an}$") predicted as one at which $R_{\min }^{2}$ vanishes, i.e., the ring
collapses to the center,
\begin{equation}
\beta _{\max }^{\mathrm{(an)}}=8\sqrt{3}\pi S.  \tag{46}
\end{equation}%
As seen in Table I, this approximate result is very close to its numerically
found counterparts at $S\geq 2$, and is quite close for $S=1$ as well.

A remarkable fact is that the analytical prediction (46) does not depend on
strength $\gamma $ of the nonlocal interaction, hence it also predicts the
onset of the collapse in the usual 2D cubic NLS equation, corresponding to $%
\gamma \rightarrow 0$:
\begin{equation}
i\frac{\partial \phi }{\partial t}=-\frac{1}{2}\nabla ^{2}\phi -\beta
\left\vert \phi \right\vert ^{2}\phi .  \tag{47}
\end{equation}%
Recall that the collapse-onset threshold for the solutions of Eq. (47) with $%
S=0$ is determined by the TS norm, $N_{\mathrm{TS}}\approx 5.85$ (Berg\'{e},
1998; Sulem and Sulem, 1999; Fibich, 2015), which, in the present notation
(taking into account normalization (39)), implies
\begin{equation}
\beta _{\max }(S=0)\approx 11.7.  \tag{48}
\end{equation}%
The analytical expression (46) is not relevant for $S=0$, but the data
displayed in Table I demonstrate that Eq. (46) offers an approximate
analytical solution for the long-standing problem of the prediction of the
collapse threshold for the VR solutions of Eq. (47). In the numerical form,
critical values which are tantamount to $\beta _{\max }(S)$ were found, for $%
1\leq S\leq 5$, by Kruglov, Logvin, and Volkov (1992). However, an
analytical approximation for them was not available prior to the results
reported by Qin, Dong, and Malomed (2016).

The independence of $\beta _{\max }(S)$ with all values of $S$ on $\gamma $
is an exact property of Eq. (38). To explain it, note that, at the final
stage of the collapse, when the shrinking VR becomes extremely narrow, the
equation for the wave function is asymptotically equivalent to the
simplified equation (47), as the nonlocal term in Eq. (38) is negligible in
this limit. Therefore, the condition for the onset of the collapse is
identical in both equations, (38) and (47). However, the difference between
them is that the 2D cubic NLS equation (47) gives rise to soliton solutions
(which are TSs, both fundamental or vortical ones), solely at $\beta =\beta
_{\max }$, and they are completely unstable. On the other hand, the
LFE-induced long-range interaction in Eq. (38) helps to create fundamental
solitons and VRs with all values of $S$ at $\beta <\beta _{\max }(S)$, and
the crucially important difference is that a part of these solution families
are \emph{stable}, see below.

\begin{table}[tbp]
\centering%
\begin{tabular}{|l|l|l|l|l|l|l|l|}
\hline
$S$ & $\beta _{\max }$ & $\beta _{\max }^{\mathrm{(an)}}$ & $\beta _{\mathrm{%
st}}$ & $S$ & $\beta _{\max }$ & $\beta _{\max }^{\mathrm{(an)}}$ & $\beta _{%
\mathrm{st}}$ \\ \hline
$0$ & $11.7$ & $\mathrm{n/a}$ & $\equiv \beta _{\max }$ & $3$ & $132.5$ & $%
130.6$ & $41$ \\ \hline
$1$ & $48.3$ & $43.5$ & $11$ & $4$ & $175.5$ & $174.1$ & $57$ \\ \hline
$2$ & $89.7$ & $87.0$ & $28$ & $5$ & $218.5$ & $217.7$ & $70$ \\ \hline
\end{tabular}%
\caption{$\protect\beta _{\max }$ and $\protect\beta _{\max }^{\mathrm{(an)}%
} $: numerically obtained and analytically predicted values of the strength
of the local nonlinearity, $\protect\beta $, up to which the 2D fundamental
solitons (with $S=0$) and VRs (vortex rings, with $S=1$ and $2$) exist. $%
\protect\beta _{\mathrm{st}}$: the numerically identified stability boundary
of the VRs. Numerical data are produced by Eqs. (35), (36) and (41) (source:
Qin, Dong, and Malomed, 2016).}
\end{table}

The stability of the solitons was systematically investigated by real-time
simulations of Eq. (38) with small random perturbations added to the
stationary solutions (furthermore, the full system of coupled equations (35)
and (36) with independent random perturbations added to components $\phi
_{\uparrow }$\ and $\phi _{\uparrow }$, was also simulated, to test the
stability against breaking the symmetry between them). The fundamental
solitons ($S=0$) are stable in their entire existence region, $\beta <\beta
_{\max }\approx 11.7$.

The simulations of the evolution of the VR families reveal an internal
stability boundary, $\beta _{\mathrm{st}}(S)<\beta _{\max }(S)$ (see Table
I), the vortices being stable at
\begin{equation}
\beta <\beta _{\mathrm{st}}(S).  \tag{49}
\end{equation}%
In the interval of $\beta _{\mathrm{st}}(S)<\beta <\beta _{\max }(0)$, the
VRs are broken by azimuthal perturbations into rotating necklace-shaped sets
of fragments, which resembles the initial stage of the instability
development of vortex solitons in usual models. However, unlike those
models, in the present case the necklace does not expand, remaining confined
under the action of the effective nonlocal interaction. Typical examples of
the stable and unstable evolution of VRs are displayed in Fig. \ref%
{fig13.6=fig163}.
\begin{figure}[tbp]
\centering\includegraphics[scale=0.8]{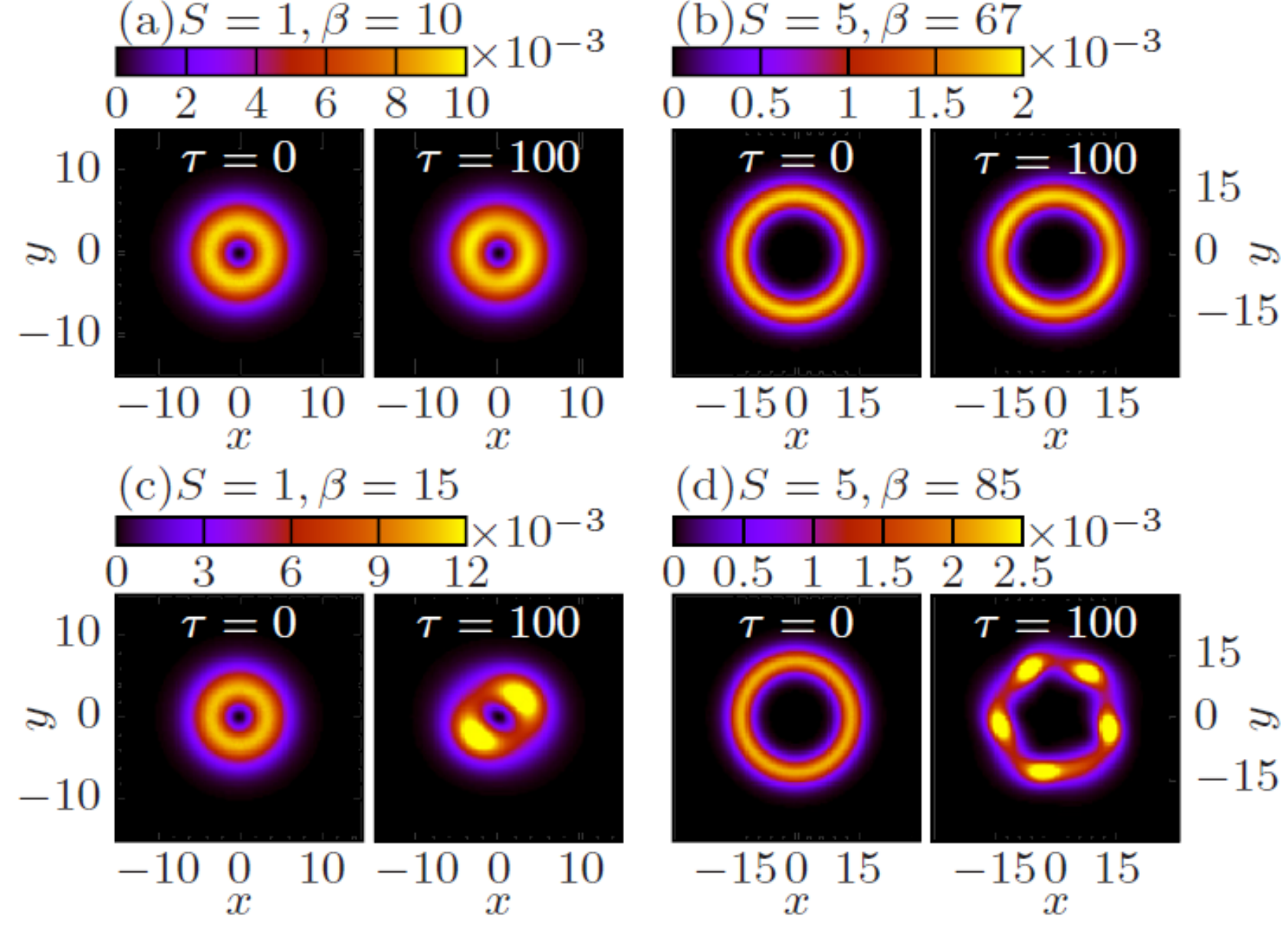}
\caption{Top and bottom panels display, severally, examples of the stable
and unstable perturbed evolution of the VRs with indicated values of $S$ and
$\protect\beta $. The initial shape of the VR ($\protect\tau =0$) is
compared to the output produced by the simulations of Eq. (38) with $\protect%
\gamma =2\protect\pi $ at $\protect\tau =100$ (here, $\protect\tau $
replaces $t$). The necklace-shaped pattern, produced by the instability in
the right bottom panel, remains confined (keeping the original overall
radius) in the course of the subsequent evolution (source: Qin, Dong, and
Malomed, 2016).}
\label{fig13.6=fig163}
\end{figure}

To address the stability of the VRs against azimuthal perturbations
analytically, one can approximate the wave function of a perturbed
(azimuthally modulated) VR by
\begin{equation}
\phi \left( r,\theta ,t\right) =A(\theta ,t)\Phi _{S}(r).  \tag{50}
\end{equation}%
This ansatz is substituted in Eq. (38), and an effective evolution equation
for the modulation amplitude $A$ is derived by averaging in the radial
direction:
\begin{equation}
i\frac{\partial A}{\partial \tau }=-\frac{1}{2R^{2}}\frac{\partial ^{2}A}{%
\partial \theta ^{2}}+\left[ \frac{\gamma \ln R}{4\pi R}-\frac{2\beta ^{2}}{%
3\left( 8\pi R\right) ^{2}}\right] |A|^{2}A.  \tag{51}
\end{equation}%
In the framework of Eq. (51), straightforward analysis of the modulational
stability of the solution with $|A|=1$ against perturbations $\sim \exp
\left( ip\theta \right) $ with integer winding numbers $p$ shows that the
stability is sustained under the threshold condition,
\begin{equation}
p^{2}\geq \left( 8/3\right) \left( \beta /8\pi \right) ^{2}.  \tag{52}
\end{equation}%
Further, the numerical results demonstrate that, similar to what is known in
other models (Quiroga-Texeiro and Michinel (1997); Pego and Warchall (2002);
Brtka, Gammal, and Malomed (2010); Reyna \textit{et al}. (2016)), the
critical instability corresponds to $p^{2}=S^{2}$ (for instance, the
appearance of five fragments in the part of Fig. \ref{fig13.6=fig163}
corresponding to $S=5,\beta =85$ demonstrates that, for $S=5$, the dominant
splitting mode indeed has $p=5$).\ Thus, the analytical prediction is that
the VRs remain stable if $p^{2}=S^{2}$ satisfies condition (52), i.e., at
\begin{equation}
\beta <\beta _{\mathrm{st}}^{(\mathrm{an})}(S)=2\sqrt{6}\pi S\approx
\allowbreak 15.4S  \tag{53}
\end{equation}%
(note that Eq. (53) does not include $\gamma $, being universal, in this
sense). On the other hand, the numerically found stability limits collected
in Table I obey an empirical formula,
\begin{equation}
\beta _{\mathrm{st}}^{(\mathrm{num})}(S)\approx \allowbreak 15S-4.  \tag{54}
\end{equation}%
The inference is that the analytical approximation (53) is quite accurate
for $S\geq 2$.

The implication of Eqs. (46), (53), and (54) is that the giant VRs, with
large values of $S$, are \emph{much more robust} than their counterparts
with smaller $S$. As mentioned above, this counter-intuitive feature is
opposite to what was previously found in those models which are able to
produce stable VRs with $S>1$ (Quiroga-Texeiro and Michinel (1997); Pego and
Warchall (2002); Borovkova \textit{et al}. (2011a,b); Driben \textit{et al}.
(2014); Sudharsan \textit{et al}. (2015); Reyna \textit{et al}. (2016)). It
is relevant to stress that, at $\beta <\beta _{\mathrm{st}}(S=0)$, the
fundamental soliton ($S=0$) is the system's ground state, while, at $\beta $
$>$ $\beta _{\mathrm{st}}(S=0)$, the ground state does not exist, due to the
possibility of the collapse. The vortices with $\beta _{\mathrm{st}%
}(S)>\beta $ cannot represent the ground state, but, nevertheless, they
exist as metastable ones, cf. the above-mentioned result for metastable 3D
solitons in the SOC system which exist while the system does not have a
ground state, due to the presence of the supercritical collapse (Zhang
\textit{et al}, 2015b).

In the case of the strong \emph{repulsive} local interaction, which
corresponds to a large positive coefficient $-\beta $ in Eq. (38), solitons
with $S=0$ can be constructed by means of the Thomas-Fermi (TF)\
approximation. In this case, instead of using the Green's function, it is
more convenient to apply the TF approximation directly to Eqs. (30) and
(31), in which the kinetic-energy terms $\sim \nabla ^{2}$ are dropped,
while $\nabla ^{2}H$ is kept in the Poisson equation (33). The corresponding
result is
\begin{equation}
\left( \Phi _{0}^{2}\right) _{\mathrm{TF}}\left( r\right) =\left\{
\begin{array}{c}
\phi _{0}^{2}J_{0}\left( \xi r\right) ~~\mathrm{at~~}r<r_{1}/\xi ,\  \\
0~~\mathrm{at}~~r>r_{1}/\xi \ ,%
\end{array}%
\right.  \tag{55}
\end{equation}%
where $\xi \equiv \sqrt{\gamma /\left\vert \beta \right\vert }$, $%
r_{1}\approx 2.4$\ is the first zero of Bessel function $J_{0}\left(
r\right) $, and $\phi _{0}$\ is a normalization constant. Figure \ref%
{fig13.7=fig164} shows that the TF approximation agrees very well with the
numerical solution.
\begin{figure}[tbp]
\centering\includegraphics[scale=0.5]{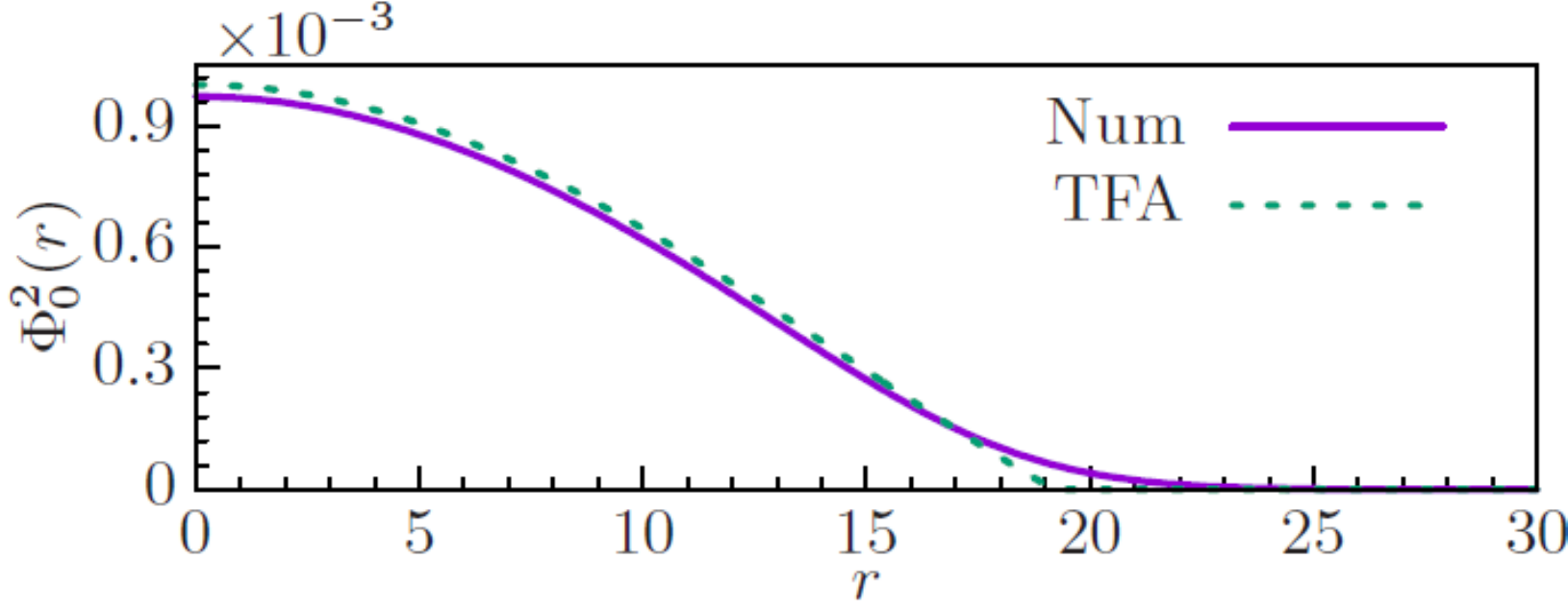}
\caption{Comparison of the TF (Thomas-Fermi) approximation, as given by Eq.
(55), for the fundamental soliton (the dashed line) and its counterpart
produced as a numerical solution of Eq. (41) the solid line), for $\protect%
\beta =-200$ and $\protect\gamma =2\protect\pi $ (source: Qin, Dong, and
Malomed, 2016).}
\label{fig13.7=fig164}
\end{figure}

\subsection{Self-accelerating 2D vortex rings (VRs)}

In the studies of diverse nonlinear-wave systems, much interest was drawn to
states which can exhibit self-accelerating and/or self-bending motion. A
well-known example is provided by Airy waves, which were predicted by Berry
and Balazs (1979) as an exact solution of the 1D linear Schr\"{o}dinger
equation. Using the similarity of the linear Schr\"{o}dinger equation to the
paraxial wave-propagation equation in classical physics, this concept was
transferred to many areas of classical and semi-classical phenomenology,
including optics (Siviloglu and Christodoulides, 2007), plasmonics (Minovich
\textit{et al}. 2013), BEC (Efremidis, Paltoglou, and von Klitzig, 2013),
acoustics (Zhang \textit{et al}., 2014), gas discharge (Clerici \textit{et al%
}., 2015), electron beams (Voloch-Bloch \textit{et al}., 2013), and studies
of surface waves in hydrodynamics (Fu \textit{et al.}, 2015).

Ideal Airy waves, with slowly decaying one-sided oscillatory tails, carry an
infinite norm, which makes them unphysical objects. To problem was resolved
by using \textit{truncated Airy waves} with a finite norm (Siviloglou and
Christodoulides, 2007; Siviloglou \textit{et al.}. 2007). However, the
truncation leads to gradual degradation of the self-accelerating wave
packets. Another caveat is that, while Airy waves are eigenmodes of linear
media, nonlinearity causes their deformation and destruction, as studied in
detail in various settings (Ellenbogen \textit{et al}., 2009; Jia \textit{et
al}., 2010; Hu \textit{et al}., 2010; Kaminer, Segev, and Christodoulides,
2011; Lotti \textit{et al}., 2011; Fattal, Rudnick, and Marom, 2011).

On the other hand, the nonlinearity of the medium, which tends to destroy
Airy waves, can be used, instead, to \emph{create} well-localized
eigenmodes, which are able to move with self-acceleration, remaining robust
objects. In addition to their stability in the presence of the nonlinearity,
an asset of such modes is that they do not develop extended tails, hence
their naturally defined norm is convergent. In particular, it was predicted
by Batz and Peschel (2013) and experimentally demonstrated by Wimmer \textit{%
et al}. (2013) that two pulses subject to the action of the group-velocity
dispersion with opposite signs may form a self-accelerating bound state in a
photonic-crystal fiber. This is possible because the pulses may be
considered as quasi-particles with opposite signs of the effective mass,
hence the opposite forces of the interaction between the pulses drive both
of them with identical signs of the acceleration. For solitons, a similar
possibility was elaborated by Sakaguchi and Malomed (2019), in a system of
nonlinearly coupled GP equations with an optical-lattice potential, where
solitons with positive and negative effective masses (it is well known that
the effective mass may be negative for gap solitons) form stable
self-accelerating pairs.

Continuing the work in this direction, it was shown by Qin \textit{et al}.
(2019) that the GP-Poisson system, based on Eqs. (30), (31) and (33), admits
\emph{exact realization} of accelerating motion of 2D solitons. In this
case, Eqs. (30) and (31) may also include the above-mentioned
self-interaction terms $\sim \tilde{\beta}$, but the Poisson equation should
not include the term $\sim k^{2}$ which is present in the Helmholtz equation
(32). Thus, the relevant system is%
\begin{equation}
i\frac{\partial \phi _{\downarrow }}{\partial t}=\left( -\frac{1}{2}\nabla
^{2}+\eta \mathbf{-}\beta \left\vert \phi _{\uparrow }\right\vert ^{2}-%
\tilde{\beta}\left\vert \phi _{\downarrow }\right\vert ^{2}\right) \phi
_{\downarrow }-\gamma H^{\ast }\phi _{\uparrow },  \tag{56}
\end{equation}%
\begin{equation}
i\frac{\partial \phi _{\uparrow }}{\partial t}=\left( -\frac{1}{2}\nabla
^{2}-\eta \mathbf{-}\beta \left\vert \phi _{\downarrow }\right\vert ^{2}-%
\tilde{\beta}\left\vert \phi _{\uparrow }\right\vert ^{2}\right) \phi
_{\uparrow }-\gamma H\phi _{\downarrow },  \tag{57}
\end{equation}%
\begin{equation}
\nabla ^{2}H=-\phi _{\downarrow }^{\ast }\phi _{\uparrow }.  \tag{58}
\end{equation}%
The system is characterized by its energy (Hamiltonian),%
\begin{equation*}
E=\int d\mathbf{r}\left[ \frac{1}{2}\left( \left\vert \nabla \phi
_{\downarrow }\right\vert ^{2}+\left\vert \nabla \phi _{\uparrow
}\right\vert \right) +\eta \left( \left\vert \phi _{\downarrow }\right\vert
^{2}-\left\vert \phi _{\uparrow }\right\vert ^{2}\right) \right.
\end{equation*}%
\begin{equation}
\left. -\beta \left\vert \phi _{\downarrow }\right\vert ^{2}\left\vert \phi
_{\uparrow }\right\vert ^{2}-\frac{\tilde{\beta}}{2}\left( \left\vert \phi
_{\downarrow }\right\vert ^{4}+\left\vert \phi _{\uparrow }\right\vert
^{4}\right) -\gamma \left( \phi _{\downarrow }^{\ast }H^{\ast }\phi
_{\uparrow }+\phi _{\downarrow }H\phi _{\uparrow }^{\ast }\right)
+\left\vert \nabla H\right\vert ^{2}\right] .  \tag{59}
\end{equation}

The result reported by Qin \textit{et al}. (2019) is that Eqs. (56)-(58) are
\emph{exactly invariant} with respect to transformation from the quiescent
reference frame into a \emph{non-inertial} one, which moves, in the 2D plane
$\left( x,y\right) $, with vectorial acceleration $\mathbf{a}=\left(
a_{x},a_{y}\right) $, combined with an additional constant velocity $\mathbf{%
V}=\left( V_{x},V_{y}\right) $. The coordinates, wave functions, and
magnetic field in the traveling frame are defined as
\begin{equation}
\left\{
\begin{array}{c}
x^{\prime } \\
y^{\prime }%
\end{array}%
\right\} =\left\{
\begin{array}{c}
x-V_{x}t-\frac{1}{2}a_{x}t^{2} \\
y-V_{y}t-\frac{1}{2}a_{y}t^{2}%
\end{array}%
\right\} ,  \tag{60}
\end{equation}%
\begin{equation}
\phi ^{\prime }\left( x^{\prime },y^{\prime },t\right) =\phi \left(
x,y,t\right) \exp \left[ -i\left( a_{x}x+a_{y}y\right) t-i\left(
V_{x}x+V_{y}y\right) +i\chi (t)\right] ,  \tag{61}
\end{equation}%
\begin{equation}
\chi (t)=\frac{1}{6}\left[ \frac{\left( V_{x}+a_{x}t\right) ^{3}}{a_{x}}+%
\frac{\left( V_{y}+a_{y}t\right) ^{3}}{a_{y}}\right] ,  \tag{62}
\end{equation}%
\begin{equation}
H^{\prime }\left( x^{\prime },y^{\prime },t\right) =H\left( x,y,t\right) -%
\frac{a_{x}}{\gamma }x-\frac{a_{y}}{\gamma }y.  \tag{63}
\end{equation}%
Actually, Eqs. (60)-(63) are a generalization of the usual Galilean boost
for the accelerating reference frame.

According to Eqs. (60) and (61), coordinates $\left( x_{c},y_{c}\right) $ of
the center of the stable 2D soliton (which may be the fundamental one of VR)
moves as
\begin{equation}
x_{c}=V_{x}t+(1/2)a_{x}t^{2},~y_{c}=V_{y}t+(1/2)a_{y}t^{2}.  \tag{64}
\end{equation}%
Equation (64) represents a curvilinear trajectory in the 2D plane: at small $%
t$, it is close to a straight line with slope $x/y=V_{x}/V_{y}$, while at $%
t\rightarrow \infty $ it is asymptotically close to a line with a different
slope, $x/y=a_{x}/a_{y}$. In particular, in the case of $a_{x}=V_{y}=0$, the
trajectory is a parabola:
\begin{equation}
y_{c}=\frac{a_{y}}{2V_{x}^{2}}x_{c}^{2}.  \tag{65}
\end{equation}

Note that the solution of the 2D Poisson equation (58) for the quiescent
soliton has the standard asymptotic form far from the region where the
source of the field is located:%
\begin{equation}
H(\mathbf{r})\approx -\frac{\gamma }{2\pi }\left( \int \phi _{\downarrow
}^{\ast }\phi _{\uparrow }d\mathbf{r}\right) \ln r,  \tag{66}
\end{equation}%
cf. Eq. (34'). The difference of the magnetic-field component (63) of the
self-accelerating 2D soliton from its quiescent counterpart (66) is the
presence of the terms linear in $x$ and $y$, which implies that the
accelerating motion is maintained by the properly built background magnetic
field.

\begin{figure}[tbp]
\subfigure[]{\includegraphics[scale=0.5]{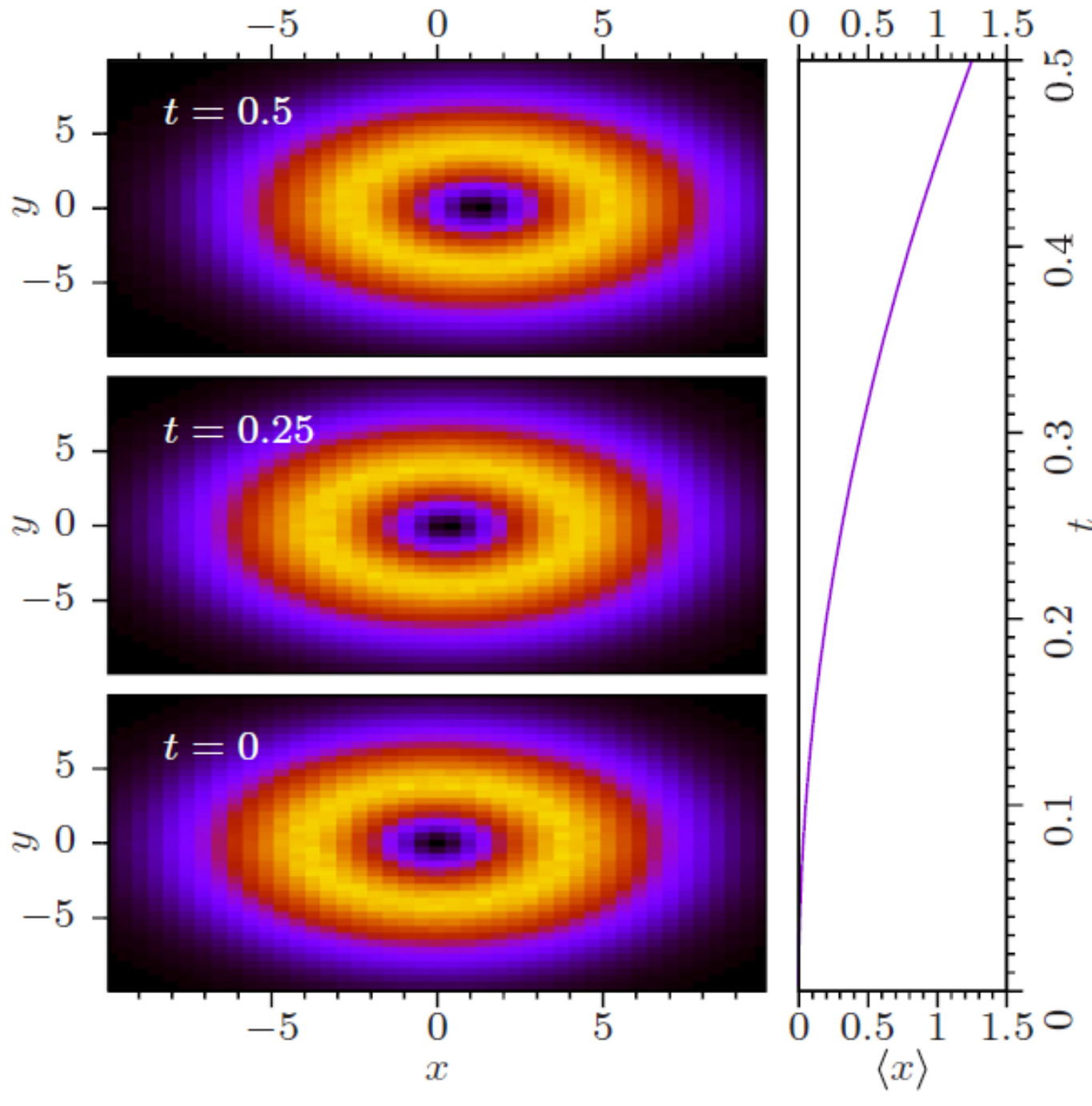}}\subfigure[]{%
\includegraphics[scale=0.5]{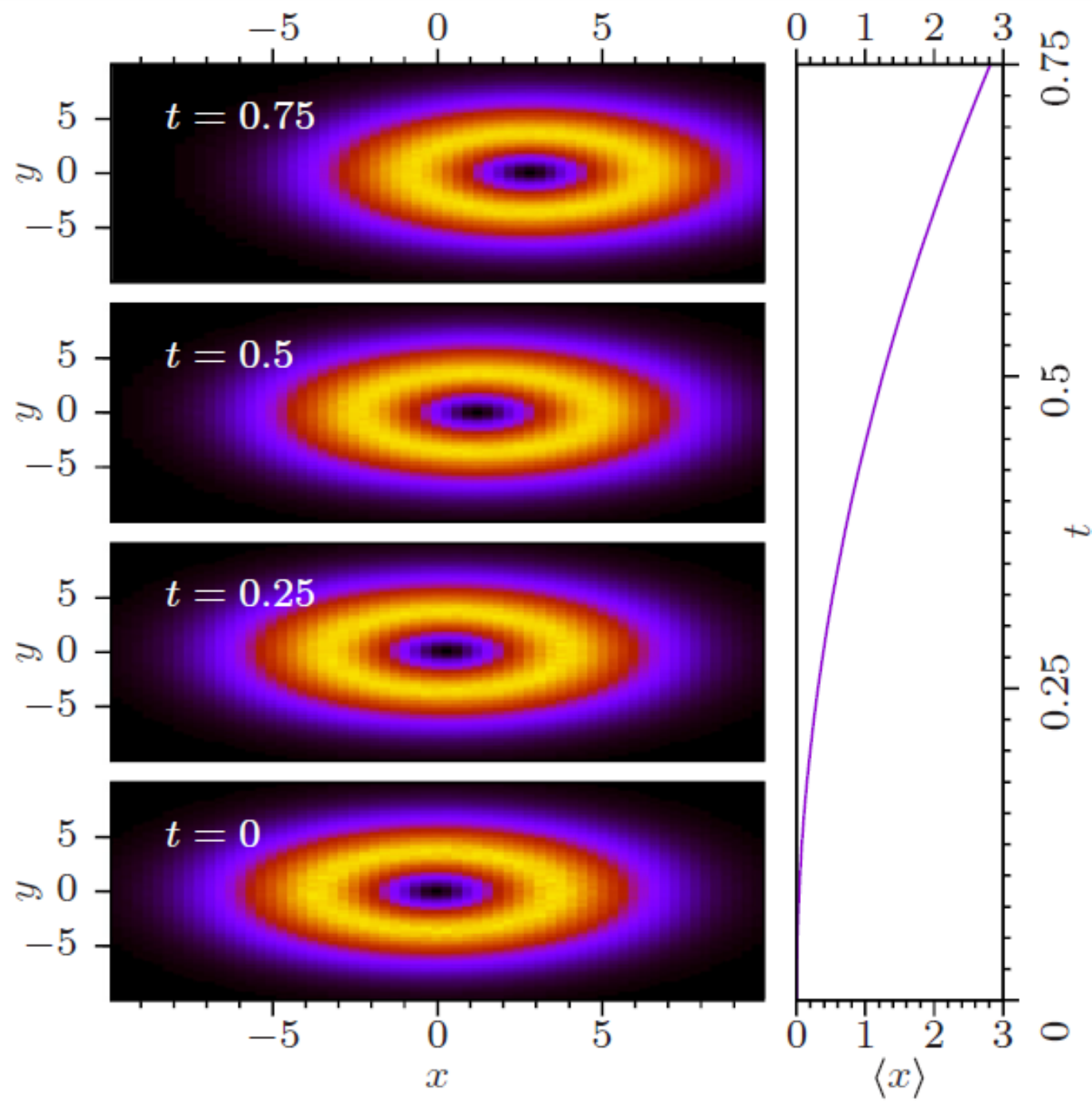}}
\caption{(a) The plot of $\left\vert \protect\phi _{\downarrow }\right\vert
^{2}=\left\vert \protect\phi _{\uparrow }\right\vert ^{2}$ for a stable
self-accelerating VR with $a_{x}=10$, $a_{y}=0$, $V_{x.y}=0$ (see Eqs. (60)
and (61)), produced by numerical solution of Eqs. (56)-(58) with $\protect%
\beta =\tilde{\protect\beta}=0$ (no local nonlinearity) and $\protect\gamma =%
\protect\pi $. The right panel shows the time dependence of coordinate $x$
of the VR's center. (b) The same as in (a), but for $\protect\beta =10$ in
Eqs. (56) and (57) (source: Qin \textit{et al.}, 2019).}
\label{fig13.8=fig166}
\end{figure}

The analytical results are corroborated by Fig. \ref{fig13.8=fig166}, which
displays numerical findings for moving VRs, produced by simulations of Eqs.
(56)-(58). These solutions demonstrate exactly the same self-accelerating
motion of the robust VRs as predicted by Eq. (65), in cases when the
underlying system does not or does contain the local nonlinear terms.

\section{Conclusion}

To keep the length of this review in reasonable limits, only a few topics
have been selected for a relatively detailed presentation. These topics are
sufficiently novel ones, while well-known results for 2D solitons stabilized
with the help of spatial nonlocality were reviewed in earlier publications
(in particular, by Krolikowski \textit{et al}., 2004; Khoo, 2009; Assanto
\textit{et al}., 2009; Peccianto and Assanto, 2012).

As concerns other aspects of this broad area, it is relevant to mention, in
particular, the theoretical and experimental results which demonstrate the
creation of stable three-dimensional QDs in single-component BECs with DDIs
between atoms carrying permanent magnetic moments (Ferrier-Barbut \textit{et
al}. 2016; Schmitt \textit{et al}. 2016; Chomaz \textit{et al}., 2016). A
possibility of the creation of QDs with embedded vorticity in this setting
was analyzed by Cidrim \textit{et al}. (2018), with a conclusion that such
states are completely unstable (on the contrary to the above-mentioned
prediction of stable 3D and 2D vortical QDs in the two-component BEC with
contact inter-atomic interactions by Kartashov \textit{et al.} (2018) and Li
\textit{et al}. (2018)).

A related finding is the prediction by Gligori\'{c} \textit{et al}. (2010)
of stable 2D solitons maintained by DDIs in a discrete system, which may be
realized by the dipolar BEC loaded into a deep optical-lattice potential. It
was also predicted by Li \textit{et al}. (2017) that the DDIs, acting along
with SOC in a two-component BEC, can maintain stable 2D gap solitons, in the
case when the kinetic energy is negligible in comparison to the SOC energy
in this system.

As for development of the work on the topic of multidimensional (chiefly,
two-dimensional) solitons in nonlocal media, an interesting direction may be
the study of interactions between such solitons, and possible formation of
their bound states. It is natural to expect that the nonlocality gives rise
to effective long-range interactions between far separated solitons, making
the situation essentially different from multidimensional settings based on
local nonlinearities, where only short-range soliton-soliton interactions
were predicted (Malomed, 1998). In particular, the long-range interactions
may affect symmetries of the emerging bound states of multidimensional
solitons. Another relevant direction is to develop the analysis for
dissipative 2D solitons in media combining nonlocal nonlinearity, gain, and
loss. Systems of this type may naturally occur in nonlinear optics.

\section*{Acknowledgments}

I would like to thank Prof. Branko Dragovi\'{c} for the invitation to submit
this paper to the special issue of journal Symmetry on the topic of
\textquotedblleft Advances in nonlinear dynamics and symmetries". This work
was supported, in part, by the Israel Science Foundation through grant No.
1286/17.

\end{document}